\documentclass[review,number,sort&compress]{elsarticle}

\usepackage{amssymb}

\usepackage{lineno}

\journal{Nucl. Instr. and Meth. Phys. Res. A}

\begin{document}

\begin{frontmatter}



\title{Energy Deconvolution of Cross Section Measurements with
an Application to the ${}^{12}{\rm C}(\alpha,\gamma){}^{16}{\rm O}$ Reaction}


\author{Carl R. Brune\corref{cor1}}
\cortext[cor1]{Corresponding author, Tel.: $+$1~740~593~1975,
  Fax: +1~740~593~1436.}
\ead{brune@ohio.edu}

\author{Daniel B. Sayre\fnref{fn1}}
\fntext[fn1]{Present address: Lawrence Livermore National Laboratory,
  Livermore, California 94550, USA}
\ead{sayre4@llnl.gov}

\address{Department of Physics and Astronomy, Ohio University,
Athens, Ohio 45701, USA}

\begin{abstract}

A general framework for deconvoluting the effects of energy averaging
on charged-particle reaction measurements is presented.
There are many potentially correct approaches to the problem; the
relative merits of some of are discussed.
These deconvolution methods are applied to recent
${}^{12}{\rm C}(\alpha,\gamma){}^{16}{\rm O}$ measurements.

\end{abstract}

\begin{keyword}
charged-particle reactions \sep computer data analysis \sep
energy deconvolution \sep
${}^{12}{\rm C}(\alpha,\gamma){}^{16}{\rm O}$



\end{keyword}

\end{frontmatter}


\section{Introduction}
\label{sec:intro}

Cross section measurements in nuclear physics in general
involve an average over a range of energies.
The importance of these convolutions depends on how strongly the
cross section varies with energy as well as experimental conditions.
In the case of experiments with charged-particle beams, which is the
intended focus of this work, the most important experimental effect
is usually beam energy loss in the target.
If these effects are significant, the data must be corrected for
them before they can be used (e.g., for astrophysical calculations,
or comparison to other experiments or theoretical calculations).
While these effects have been known by workers in the field for
many decades, a general treatment has not been given
and some confusion remains in the literature.
In a recent paper, \citet{Lem08} has advocated a particular approach
for correcting experimental data for energy averaging.
It is, however, but one of many valid approaches, and, in some circumstances
it has some distinct disadvantages.
In this paper, we present a general framework for deconvoluting the
effects of energy averaging. We emphasize that there can be
many correct approaches and discuss some of their advantages and
disadvantages.

The ${}^{12}{\rm C}(\alpha,\gamma){}^{16}{\rm O}$ reaction is a very
important process in nuclear astrophysics. 
Differential cross section data for this reaction have been recently
published by \citet{Ass06}.
This data set provides an opportunity to demonstrate energy deconvolution,
which also renders the data in a form that is more useful for
further future analysis.

\section{Energy Deconvolution}

\subsection{Statement of the Problem}

For an ideal experiment, with a mono-isotopic target of uniform areal density
and with perfect energy resolution, the reaction yield $Y$
(number of reactions per incident particle) is related to the
cross section $\sigma$ via
\begin{equation}
Y(E)=\sigma(E)\,n\,\Delta x,
\label{eq:y_zero_thick}
\end{equation}
where $n$ is the number density of target atoms, $\Delta x$ is the linear
target thickness, and $E$ is the beam energy\footnote{%
We utilize laboratory energies throughout this paper,
unless otherwise indicated.}.
In a real experiment, a variety of additional physical effects must
often be taken into consideration.
For experiments performed with charged-particle beams, which are the focus
of this work, beam energy loss in the target is an important effect.
Considering energy loss in the target,
the relationship between yield and cross section becomes
\begin{equation}
Y(E_0) = \int_0^{\Delta x} \sigma(E(x))\,n\,{\rm d}x =
  \int_{E_0-\Delta E}^{E_0} \frac{\sigma (E)}{\epsilon (E)}\,{\rm d}E,
\label{eq:y_thick}
\end{equation}
where $E_0$ is the incident beam energy,
$x$ measures the linear depth in the target,
$\Delta E$ is the energy loss in the target, and
$\epsilon=-\frac{{\rm d}E}{n{\rm d}x}$ is the stopping power.
The stopping power is energy-dependent, and $\Delta E$ must generally be
determined by a numerical procedure. For example, the variation
of the energy with depth in the target can be determined by numerically
solving the differential equation
\begin{equation}
\frac{{\rm d}E}{{\rm d}x}(x) = -n\,\epsilon (E),
\end{equation}
subject to $E(0)=E_0$, and then $\Delta E = E_0-E(\Delta x)$.
The important point that we are concerned with here is that now the
yield is related to an energy convolution of the cross section.
Energy loss in the target is usually the most important effect contributing
to the overall energy resolution in an experiment, but additional
effects such as energy straggling, the incident beam energy resolution,
energy-dependent detector efficiency, and target composition
may also be important. These additional effects lead to a more complicated
form of Eq.~(\ref{eq:y_thick}) that in some cases is most practically
implemented via a Monte Carlo simulation.
Since the nature of these effects depends upon the details of the particular
experiment, we will not consider them further here.
It should be noted that considerations discussed below can be generalized in a
straightforward manner to these more complicated energy convolutions.
Equation~(\ref{eq:y_thick}) can also be applied to differential cross section
measurements, in which case $Y$ becomes the yield per steradian and
$\sigma$ must be replaced by the differential cross section.

The question we want to address here is: how does one convert $Y(E_0)$, the
yield measured in an experiment, to a cross section? In other words,
how does one deconvolute the effects of the energy averaging process?
Our primary interest is to compare the relative merits of various
mathematical approaches that have been posed in the literature.
In order to carry this out, it is necessary that the energy dependence
of the cross section and the physical processes generating the energy
convolution be well understood.
We will, for now, assume these assumptions are fulfilled.
The important issue of error analysis arising from uncertainties in
the experimental data and the energy convolution is dependent upon
the details of the particular experiment and will be discussed briefly
at the end of this section.
The effects of angular convolution, resulting from the acceptance of
the detection system, are not considered in the analysis presented
in this section.

These deconvolution corrections are expected to be most important when
the cross section is highly energy dependent (e.g., charged-particle
reactions below the Coulomb barrier or resonances) and/or when when
thicker targets are utilized.
It should also be noted that the methods discussed in this paper
may not be the most appropriate for analyzing data from narrow resonances
(with widths significantly smaller than $\Delta E$);
in those cases one generally determines the resonance parameters
(e.g., resonance strengths and resonance energies, rather than cross sections)
directly from the measured yields.

\subsection{The Effective Energy Approach}
\label{subsec:effective}

One approach is to to utilize the ideal formula, Eq.~(\ref{eq:y_zero_thick}),
to define the experimental cross section
\begin{equation}
\sigma_{\rm exp}(E_{\rm eff}) = \frac{Y(E_0)}{n\, \Delta x},
\label{eq:y_e_eff}
\end{equation}
where $E_{\rm eff}$ is the {\em effective energy}, defined in this work
to be given implicitly by
\begin{equation}
\sigma(E_{\rm eff})\int_{E_0-\Delta E}^{E_0}\frac{{\rm d}E}{\epsilon(E)}=
  \int_{E_0-\Delta E}^{E_0} \frac{\sigma (E)}{\epsilon (E)}\,{\rm d}E.
\label{eq:e_eff}
\end{equation}
In Eq.~(\ref{eq:e_eff}), the assumed theoretical function for $\sigma(E)$
is used on both sides of the expression; the solution for $E_{\rm eff}$
thus requires inverting the cross section function.
Since this approach ensures that the measured yield and experimental cross
section are consistent with Eq.~(\ref{eq:y_thick}), this procedure
can be considered a faithful (mathematically-correct) deconvolution.
As far as we are aware, this approach was first described by
Dyer and Barnes in 1974~\cite{Dye74}, and has been used by several authors
subsequently (see, e.g., Refs.~\cite{Fil83,Imb05}).
It has also been described and advocated for in a recent paper
by \citet{Lem08}.

\subsection{A General Class of Approaches}

A more general class of deconvolution procedures can be derived by considering
adjustments to the experimental cross sections in addition
to the energy:
\begin{equation}
\sigma_{\rm exp}(\tilde{E}) = \frac{Y(E_0)}{f\,n\, \Delta x}.
\label{eq:y_f}
\end{equation}
Here, $\tilde{E}$ is the energy assigned to the experimental measurement
and the correction factor $f$ is given by
\begin{equation}
f=\frac{ \int_{E_0-\Delta E}^{E_0} \sigma(E)\,[\epsilon(E)]^{-1}\, {\rm d}E }
  {\sigma(\tilde{E})\int_{E_0-\Delta E}^{E_0} [\epsilon(E)]^{-1}\, {\rm d}E }.
\label{eq:f_factor}
\end{equation}
The factor $f$ depends upon the assumed energy dependence of $\sigma(E)$
and again the procedure
is consistent with Eq.~(\ref{eq:y_thick}), ensuring a faithful deconvolution.
The method for determining $\tilde{E}$ is not specified at this stage, but
one expects reasonable procedures to provide $\tilde{E}$ within the
range of energies in the target, i.e., $E_0-\Delta E\le\tilde{E}\le E_0$.
One can consider the effective energy approach described
in Subsec.~\ref{subsec:effective} to be a special case
where $\tilde{E}$ is adjusted to make $f=1$.

An intuitive choice for $\tilde{E}$ is to define it to be the {\em mean}
or cross-section weighted energy:
\begin{equation}
\bar{E}=\frac{ \int_{E_0-\Delta E}^{E_0} E\, \sigma(E)\,[\epsilon(E)]^{-1}\,
  {\rm d}E }
  { \int_{E_0-\Delta E}^{E_0} \sigma(E)\,[\epsilon(E)]^{-1}\, {\rm d}E }.
\label{eq:ebar}
\end{equation}
To the best of our knowledge,
this procedure was first used by \citet{Wre94}. A similar
approach is described by \citet{Dwa74}.
This choice ensures in some sense that the energy assigned
to an experimental cross section measurement reflects the most
important reaction energies in the target.
An additional utility of this method is that the mean energy can be measured
directly in some cases, via the detection of
capture gamma rays~\cite{Ale84,Bru94,Kin94}.
This approach was mischaracterized in Ref.~\cite{Lem08}, as the correction
factor $f$ was left out of that author's description.

Another possible choice for $\tilde{E}$ is to define it be the {\em median}
energy $E_{\rm m}$, which is defined implicitly:
\begin{equation}
\int_{E_0-\Delta E}^{E_0} \frac{\sigma (E)}{\epsilon (E)}\,{\rm d}E =
  2\int_{E_{\rm m}}^{E_0} \frac{\sigma (E)}{\epsilon (E)}\,{\rm d}E.
\label{eq:e_median}
\end{equation}
This definition has been given in Refs.~\cite{Rol87,Rol88,Ili07},
but these authors have wrongly implied that this median energy can
be used together with Eq.~(\ref{eq:y_e_eff}) to faithfully deconvolute
yields into cross sections. From the discussion above, it is clear
that the median energy is of no particular significance in the
deconvolution process and that Eq.~(\ref{eq:y_f}), which includes
the deconvolution factor $f$, must be used to faithfully deconvolute
yields into cross sections.
This approach was recently used by \citet{Mak09} for the analysis of
${}^{12}{\rm C}(\alpha,\gamma){}^{16}{\rm O}$ measurements.
In this case, the deconvolution factor is significant and is taken
into account, with $f_m\approx 0.77$ for both energy-target
combinations reported.

\subsection{Analytic Approximations}

We will briefly consider here some analytic approximations to the
deconvolution approaches given above. We ignore the energy dependence
of the stopping power and assume the energy dependence of the
cross section to be at most quadratic. Expanding the quadratic
around the energy at the center of the target, $E_h=E_0-\Delta E/2$, we have
\begin{equation}
\sigma(E)=\sigma_h+\left(\frac{{\rm d}\sigma}{{\rm d}E}\right)_{E_h}(E-E_h)+
  \left(\frac{{\rm d}^2\sigma}{{\rm d}E^2}\right)_{E_h}(E-E_h)^2,
\end{equation}
where $\sigma_h=\sigma(E_h)$ and the first and second derivatives are
also evaluated at $E_h$. It is also useful to define
\begin{equation}
R_1\equiv \frac{1}{\sigma_h}\left(\frac{{\rm d}\sigma}{{\rm d}E}\right)_{E_h}
  \quad {\rm and} \quad
R_2\equiv \left[ \left(\frac{{\rm d}^2\sigma}{{\rm d}E^2} \right) /
  \left(\frac{{\rm d}\sigma}{{\rm d}E}\right) \right]_{E_h}.
\end{equation}

For the case of the effective energy, Eq.~(\ref{eq:e_eff}) leads to a quadratic
equation:
\begin{equation}
R_2\varepsilon^2 +2\varepsilon-\frac{1}{12}R_2(\Delta E)^2=0,
\label{eq:eff_quad}
\end{equation}
where $E_{\rm eff}=E_0-\Delta E/2 +\varepsilon$.
The resulting effective energy is
\begin{equation}
E_{\rm eff}=E_0-\frac{1}{2}\Delta E +
  \frac{[1+R_2^2(\Delta E)^2/12]^{1/2}-1}{R_2},
\end{equation}
where we have adopted the solution which approaches $E_0$ in the
$\Delta E\rightarrow 0$ limit. This point will be discussed further below.
It is interesting to note that if $R_2=0$, i.e., if the cross section
depends only linearly upon the energy, then using
$E_{\rm eff}=E_0-\Delta E/2$ provides an exact deconvolution
(the energy dependence of the stopping power is also ignored).

For the case of the mean energy, we set $R_2=0$, due to the more
complicated integrations required. The results are
\begin{equation}
\bar{E}=E_0-\frac{1}{2}\Delta E +\frac{1}{12}R_1(\Delta E)^2
\end{equation}
and
\begin{equation}
f^{-1} = 1+\frac{1}{12}R_1^2(\Delta E)^2.
\end{equation}

For the the case of the median energy, we also take $R_2=0$, resulting in
\begin{equation}
E_{\rm m}=E_0-\frac{1}{2}\Delta E+\frac{[1+R_1^2(\Delta E)^2/4]^{1/2}-1}{R_1}
\label{eq:med_lin}
\end{equation}
and
\begin{equation}
f_{\rm m}^{-1}=\left[1+\frac{1}{4}R_1^2(\Delta E)^2\right]^{1/2}.
\end{equation}
Equation~(\ref{eq:med_lin}) is mathematically equivalent to Eq.~(5.55) of
Ref.~\cite{Rol88}, but, again, that reference did not consider
the need for the deconvolution factor $f_m$.

\subsection{An Example Case}

\begin{figure}[tbh]
\begin{center}
\includegraphics[width=4in]{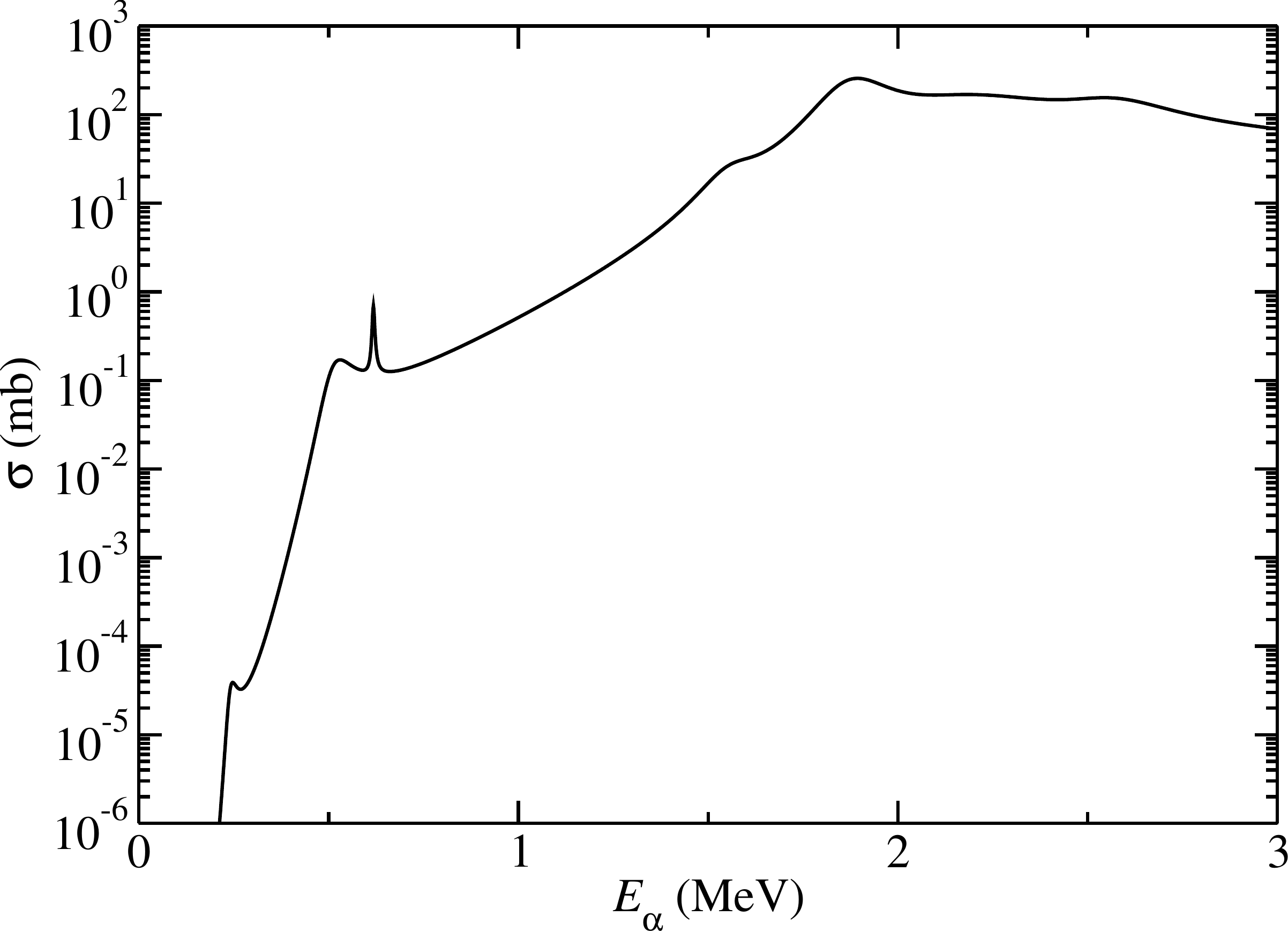}
\caption{The ${}^9{\rm Be}(\alpha,n)$ total cross section, as a function
of laboratory $\alpha$ energy, calculated
from the parameters given in Table~I of Ref.~\protect\cite{Wre94}.}
\label{fig:be9an_sig}
\end{center}
\end{figure}

A good example of the application of these methods is provided
by the ${}^9{\rm Be}(\alpha,n)$ reaction, which was measured for
$0.23\le E_\alpha\le2.7$~MeV by~\citet{Wre94}.
Over this energy range, the cross section includes the effects
of the Coulomb barrier as well as narrow and broad resonances.
The cross section varies by eight orders of magnitude
over this range and is shown in Fig.~\ref{fig:be9an_sig}.
Three targets were utilized for the measurements reported in Ref.~\cite{Wre94},
consisting of $5.08\times 10^{16}$ (target 1),
$1.23\times 10^{18}$ (target 2), and $3.23\times 10^{18}$ (target 3)
atoms/cm${}^2$ of ${}^9{\rm Be}$, respectively.
These data were analyzed using the mean energy method described
above, with the energy dependence of the detector efficiency also
taken into consideration. Consistent results for the cross section were
deduced near the 0.618-MeV resonance using all three targets, and for energies
down to $E_\alpha\approx 0.22$~MeV using targets~2 and~3.

\begin{figure}[tbh]
\begin{center}
\includegraphics[width=4in]{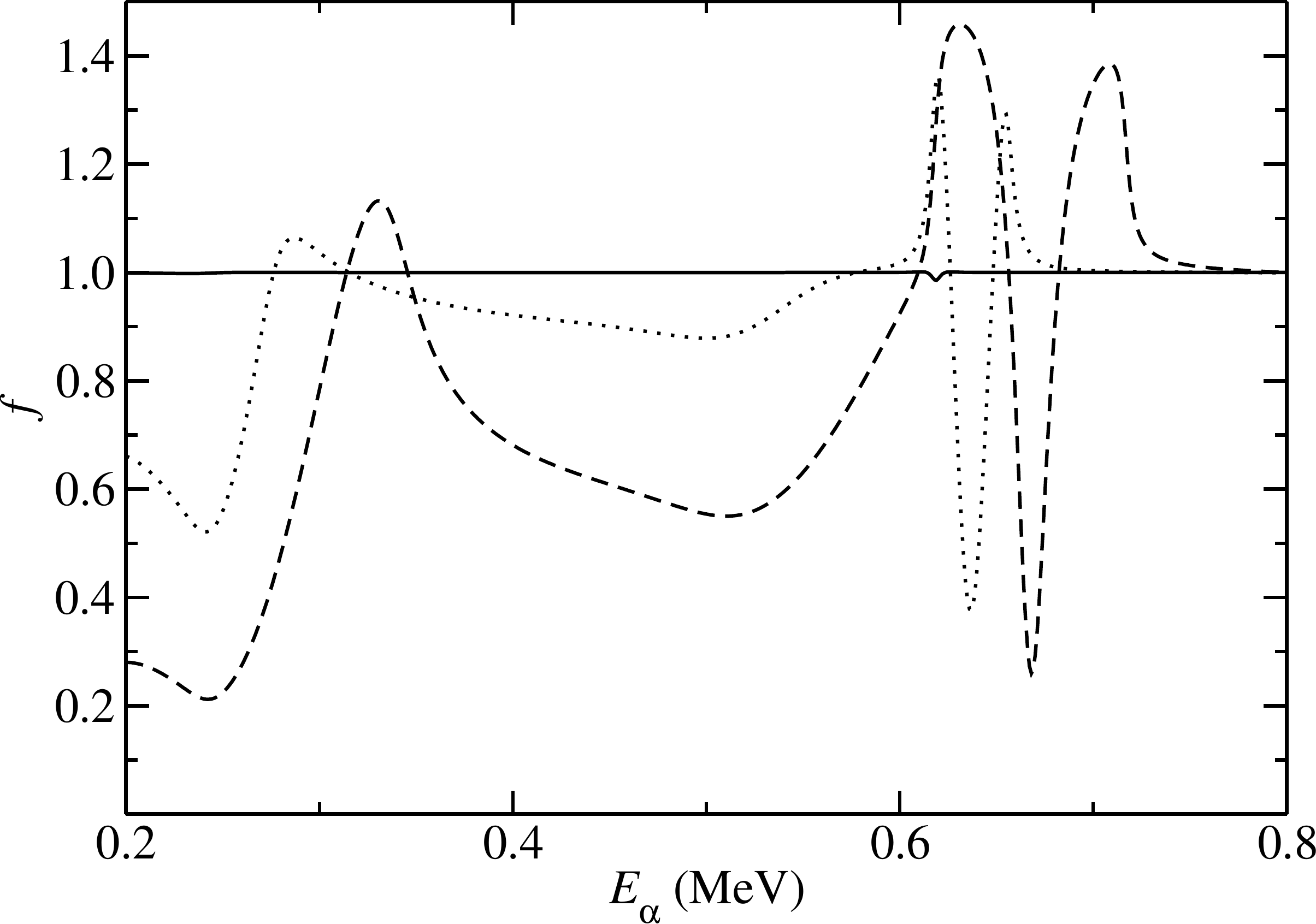}
\caption{The correction factor $f$ for target~1 (solid curve),
target~2 (dotted curve), and target~3 (dashed curve).}
\label{fig:be9an_f}
\end{center}
\end{figure}

Using the energy dependence of the cross section defined by the
parameters in Table~I of Ref.~\cite{Wre94} and also the stopping
power from that reference, we have calculated
the mean energies and corresponding correction factors for these targets
using Eqs.~(\ref{eq:f_factor}) and~(\ref{eq:ebar});
we have ignored the beam-energy dependence of the detection
that was also taken into consideration by Ref.~\cite{Wre94}.
The resulting correction factors are shown in Fig.~\ref{fig:be9an_f}
for low energies. As expected, the correction deviates more from unity
for thicker targets and/or when there is a strong energy dependence of
the cross section due to resonances or the Coulomb barrier.
The results using the median energy are qualitatively similar and
will not be discussed further.
For the effective energy approach, $f\equiv 1$ and all the deconvolution
is contained in the effective energy.

\subsection{Discussion}

\begin{figure}[tbhp]
\begin{center}
\includegraphics[width=4in]{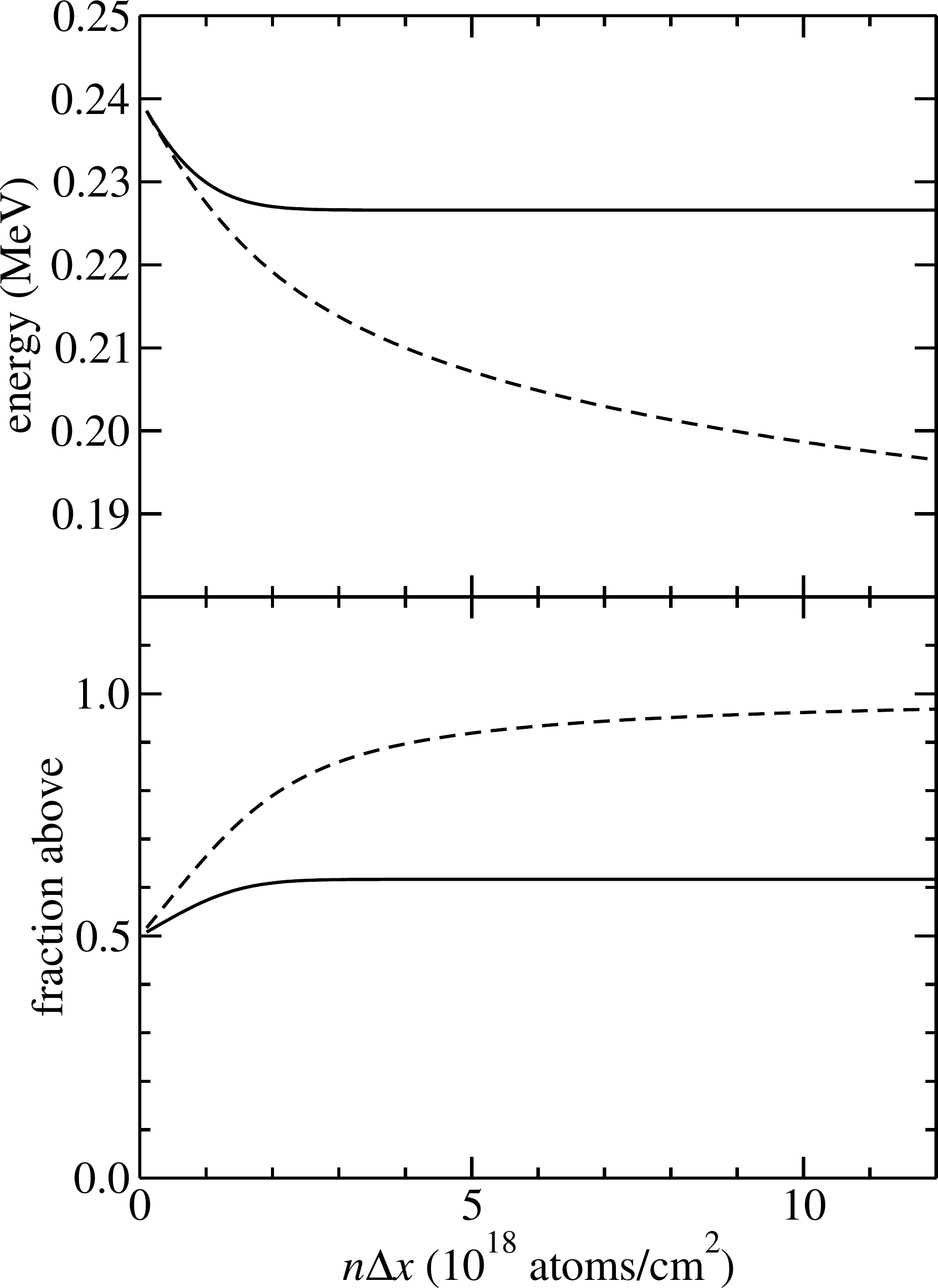}
\caption{The dependence of the mean energy (solid curve) and effective
energy (dashed curve) on the target thickness is shown in the upper panel,
for ${}^9{\rm Be}(\alpha,n)$ and an incident $\alpha$ energy of 0.24~MeV.
The lower panel shows the corresponding fractions of the yield resulting
from reaction energies above the mean energy (solid curve) and
above the effective energy (dashed curve).}
\label{fig:be9an_nx}
\end{center}
\end{figure}

It is found that there are many possible approaches to the energy
deconvolution that are mathematically exact. However, in practice,
these different choices may lead to different results, due to the fact that
the energy dependence of the cross section and the nature of the
energy convolution are not exactly known. It may also be the case in
practice that approximate deconvolutions, such as ignoring the
correction factor $f$, may provide perfectly acceptable results.

In the preliminary stages of the analysis of the data presented in
Ref.~\cite{Wre94}, the effective energy approach described above was
attempted. Two shortcomings of this method were encountered. The first
problem is related to the fact that it requires inverting the cross
section function. If the cross section has a local maxima or minima
(e.g., at the peaks of or between resonances), this inversion is ambiguous.
A criterion for selecting the solution must then be adopted;
choosing the one nearest $E_0-\Delta E/2$ is one reasonable convention.
In any case, one finds that $E_{\rm eff}(E_0)$ is a discontinuous function
and that certain ranges of $E_{\rm eff}$ cannot be reached for any $E_0$.
Effectively, data points are pushed away in energy from
cross section maxima or minima. 
The analytic approximation for the effective energy given by
Eq.~(\ref{eq:eff_quad}) is a quadratic equation and provides
an example of this ambiguity.
If the cross section has a local extrema at the center of the target,
i.e., if $({\rm d}\sigma/{\rm d}E)_{E_h}= 0$, then
$E_{\rm eff}=E_0-\Delta E/2\pm\Delta E/\sqrt{12}$,
and the choice of roots in ambiguous.
In the case of resonances (maxima) this behavior was not deemed
desirable because the presence of the resonance was largely responsible
for the measured yield and one expects reasonable procedures to assign an
energy to a measurement that reflects the reaction energies that
gave rise to the measured yield.

Another shortcoming noted is that for thicker targets at energies
well below the Coulomb barrier, $E_{\rm eff}$ is located at an energy
significantly less than the energies where the majority of the
yield is generated.
In Fig.~\ref{fig:be9an_nx}, the mean and effective energies are plotted
versus the target thickness, for the ${}^9{\rm Be}(\alpha,n)$ reaction
with an indicant $\alpha$ energy of 0.24~MeV (approximately the lowest
energy measured in Ref.~\cite{Wre94}).
It is seen that the mean energy approaches a constant value once a
thickness of $\approx 2\times 10^{18}$/cm${}^2$ has been reached,
as additional target thickness does not contribute to the
integrals in Eq.~(\ref{eq:ebar}).
However, the effective energy continues to decrease as the target thickness
is increased, as additional target thickness does continue to contribute
to the integral on the left-hand side of Eq.~(\ref{eq:e_eff}).
The lower panel of Fig.~\ref{fig:be9an_nx} shows the fraction 
of the yield resulting from reaction energies above mean and effective
energies. It is seen that for the mean energy the fraction remains
nearly one half, while for the case of the effective energy it approaches one.
This behavior of the effective energy is not desirable, as the effective
energy provides a misleading picture of what reaction energies were
actually measured. In this situation the effective energy approach will also
be more prone to systematic errors resulting from imprecise knowledge
of the energy dependence of the cross section.

It was for the above reasons that the effective energy was not
used by \citet{Wre94}. Another situation where the effective energy
approach is less useful is when multiple yields are analyzed from
a single bombardment, such as for cross sections to different final states
or differential cross sections for different angles.
The energy dependences of these cross sections may be quite different,
e.g., because of resonances involving particular partial waves,
resulting in different effective energies being assigned to the
different measurements. The resulting data are much more difficult to present
and further analyze: consider, for example, an angular distribution where
each angle corresponded to a different energy!
This complication can be avoided via the mean or median energy
approaches, using the total cross section to calculate the mean
or median energy, and then applying the appropriate
correction factors to the individual partial or differential
cross sections. This approach is used below for the analysis of 
${}^{12}{\rm C}(\alpha,\gamma){}^{16}{\rm O}$ angular distribution data.

The above difficulties with the effective energy method are
related to the fact that all of the deconvolution is performed by
adjusting the effective energy.
These difficulties can be avoided by adopting alternative choices for
the energy and determining the experimental cross section via
Eqs.~(\ref{eq:y_f}) and~(\ref{eq:f_factor}).
Both the mean and median energies, defined by Eqs.~(\ref{eq:ebar})
and~(\ref{eq:e_median}), are reasonable choices that assign energies
to the experimental cross section that accurately reflect
the energies that were measured.
We are not aware of any cases where there would be significant differences
between these two approaches.
We have a slight preference for the mean energy, as it is easier
to calculate than the other energy definitions discussed here.

In all of this discussion, it has been assumed that the convolution kernel
and the energy dependence of the cross section are known.
In practice these inputs may not be precisely determined, and error
analysis must be considered.
For  the cross section extraction, it has been found that using
self-consistent iterative procedures to determine the energy
dependence of the cross section work quite well in several
cases; see, e.g., \citet{Wre94,Lem08}.
Uncertainties arising from the cross section assumptions and the
possibility missing features in the cross section due to insufficient
energy resolution must be considered.
Particular attention should be paid to the lowest-energy cross section
data point (or points), where the assumed energy dependence of the
cross section is obviously not constrained by any lower-energy data points.
In addition, the energy regions above strong resonances should be
scrutinized for the possibility that a tail in the energy resolution
function causes measurements above the resonances to receive
resonant contributions. The latter effect is important in the
case discussed in the following section.

\section{Application to ${}^{12}{\rm C}(\alpha,\gamma){}^{16}{\rm O}$ data}

Extensive angular distribution measurements of the
${}^{12}{\rm C}(\alpha,\gamma_0){}^{16}{\rm O}$ reaction
were performed at Stuttgart a few years ago using detectors
from the EUROGAM array~\cite{Ass06,Fey04}.
Measurements of this type are necessarily taken with targets which
give rise to significant energy averaging effects.
The energy deconvolution is complicated by the fact that the
energy dependence of the differential cross section depends upon the angle,
particularly near the $E_{c.m.}=2.68$-MeV resonance.
In Refs.~\cite{Ass06,Fey04}, a complete energy deconvolution was not attempted.
The primary goal of the present analysis is to more fully take into account
the effects of energy averaging and to put the data into a form that
is useful for further analyses, including a simultaneous $R$-matrix
analysis of the angular distributions.
New information about the interference structure involving the
$E_{c.m.}=2.68$-MeV resonance~\cite{Say12} is taken into account.

\subsection{Cross Section Model}

\begin{figure}[tbhp]
\begin{center}
\includegraphics[width=4in]{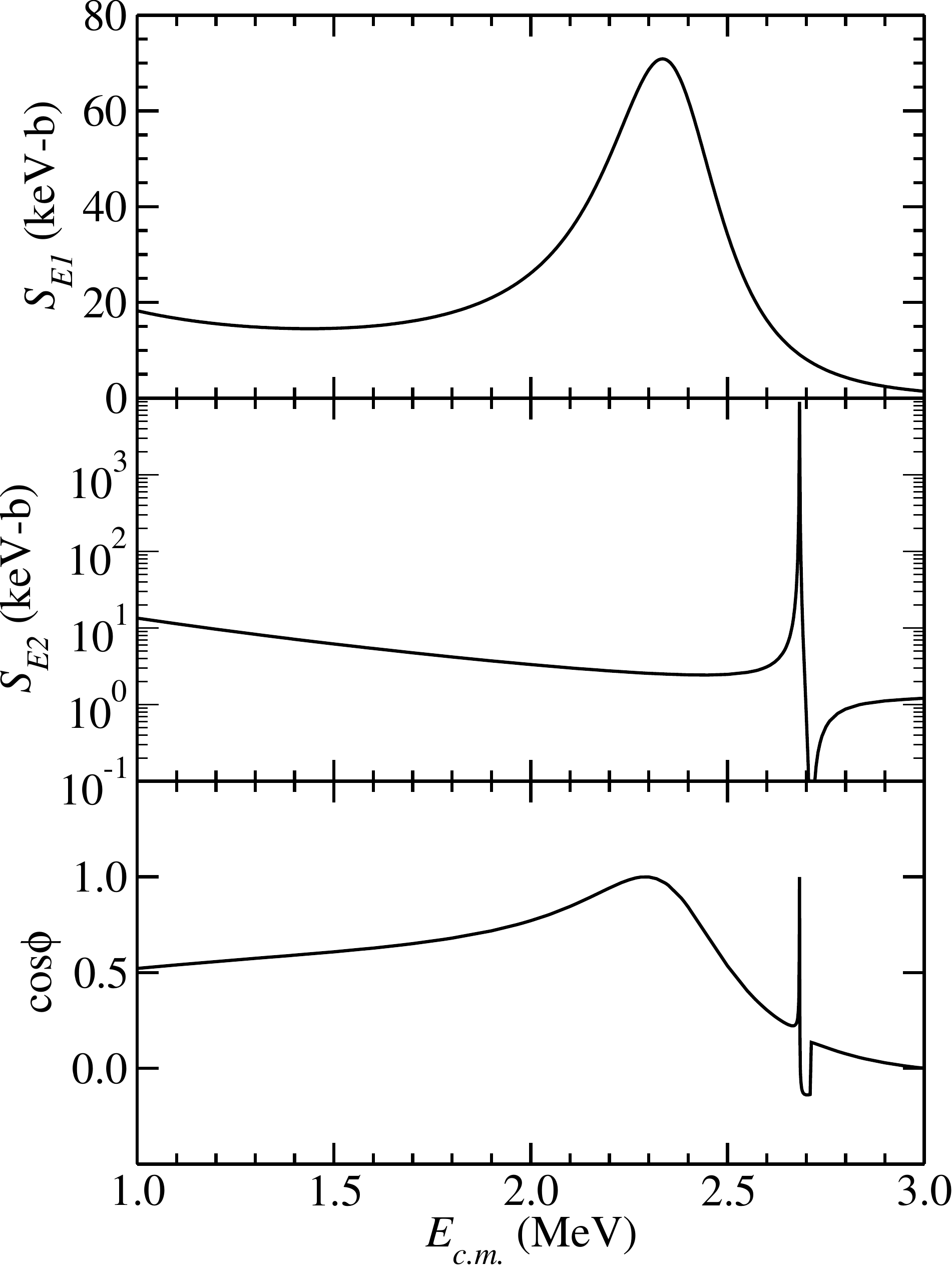}
\caption{The $R$-matrix parametrizations for $S_{E1}$, $S_{E2}$,
and $\cos\phi$ as a function of CM energy.}
\label{fig:rmatrix}
\end{center}
\end{figure}

In order to consider the effects of energy convolution on the
differential cross section, a model for its energy dependence must be adopted.
The differential cross section for
${}^{12}{\rm C}(\alpha,\gamma_0){}^{16}{\rm O}$
can then be written as~\cite{Dye74,Bar91}
\begin{eqnarray}
\frac{{\rm d}\sigma}{{\rm d}\Omega}&=&\frac{1}{4\pi}\biggl[ \sigma_{E1}(1-P_2)
+\sigma_{E2}(1+\frac{5}{7}P_2-\frac{12}{7}P_4) \nonumber \\
&&+6\cos\phi\sqrt{\frac{\sigma_{E1}\sigma_{E2}}{5}}(P_1-P_3)\biggr],
\label{eq:dsdo}
\end{eqnarray}
where $\sigma_{E1}$ and $\sigma_{E2}$ are the $E1$ and $E2$ ground-state
cross sections, $\phi$ is the relative phase, and the angular dependence is
contained in the Legendre polynomials $P_l\equiv P_l(\cos\theta)$.
The term $\propto\sqrt{\sigma_{E1}\sigma_{E2}}$ results from $E1$-$E2$
interference, and gives rise to an asymmetry around $90^\circ$
in the differential cross section.
It is often convenient to to use the $S$-factors
$S_{EL}=\sigma_{EL}E\exp(2\pi\eta)$, where $\eta$ is the Coulomb parameter.
Equation~(\ref{eq:dsdo}) can also be written as
\begin{equation}
\frac{{\rm d}\sigma}{{\rm d}\Omega}=\frac{\sigma_0}{4\pi}W(\theta),
\quad {\rm with} \quad W(\theta)=1+\sum_{l=1}^4 a_l P_l,
\label{eq:dsdo_w}
\end{equation}
where
$\sigma_0=\sigma_{E1}+\sigma_{E2}$,
\begin{eqnarray}
a_1 = -a_3 &=& \frac{6\cos\phi}{\sigma_0}
  \sqrt{\frac{\sigma_{E1}\sigma_{E2}}{5}}, \\
a_2 &=& \frac{-7\sigma_{E1}+5\sigma_{E2}}{7\sigma_0}, {\rm ~and} \\
a_4 &=& -\frac{12}{7}\frac{\sigma_{E2}}{\sigma_0}.
\end{eqnarray}
If the above theoretical expressions for the differential cross section
are to be compared to experimental differential cross sections
measured with a finite solid angle, $P_l$ should be replaced with
$Q_l P_l$, where the $Q_l$ are the attenuation
factors~\cite{Ros53}.
We will utilize $W_Q(\theta)$ and $W(\theta)$ to indicate angular
dependences with and without the attenuation coefficients, respectively.

For computing the convolutions, we assume that $\sigma_{E1}$,
$\sigma_{E2}$, and $\cos\phi$ are given by $R$-matrix parametrizations,
as shown in Fig.~\ref{fig:rmatrix}, where
for convenience $S$ factors are shown in place of cross sections.
The $E1$ parametrization is essentially the same as that described
in Ref.~\cite{Bru99}, except that more recent phase shift data~\cite{Tis09}
has been fitted. The $E2$ parametrization is taken from
Ref.~\cite{Say12} (the solid curve in the lower panel of Fig.~2).
The parametrization for $\cos\phi$ is taken from the relative phase
of the above $E1$ and $E2$ parametrizations.
It should be noted that $\cos\phi$ is primarily defined by the
P- and D-wave scattering phase shifts~\cite{Bru01} and that our
parametrization accurately reproduces the recently-published
phase shifts of \citet{Tis09}.
Many of the details of these parametrizations are unimportant for the
convolutions discussed below.
The essential ingredients are the roughly constant $S$-factors below
$E_{c.m.}=2$~MeV, and the behavior of $\sigma_{E1}/\sigma_{E2}$ and $\cos\phi$
near the narrow $E_{c.m.}=2.68$-MeV resonance.
This behavior is largely determined by the known resonance
parameters, but the nature of the interference was until
recently not understood~\cite{Bru01}.
However, the experiment and analysis of \citet{Say12} have now conclusively
determined how this resonance interferes with other $E1$ and $E2$ amplitudes.

\subsection{The Experimental Data}
\label{subsec:fey_data}

Refs.~\cite{Ass06,Fey04} report measurements at 25 bombarding energies
that are described in Tables~I and~II of \citet{Ass06};
particular energies will be referred to by measurement numbers 1-25
which are ordered by energy as given in these tables.
Gamma-ray angular distributions were measured at nine angles and
are given in Figs.~D.1-9 of \citet{Fey04}.
These data are the efficiency-corrected numbers of gamma rays detected,
scaled by $10^{-4}$; these quantities are defined
here to be $N_\gamma(\theta)$.
The data have also been corrected for finite solid angle.
When normalized by the factor $0.624(4\pi)Q\tilde{N}$, where $Q$ and
$\tilde{N}$ are given in Table~II of Ref.~\cite{Ass06}, these data
become differential cross sections in nb/sr.
We have verified that these differential cross sections, when fitted with
Eq.~(\ref{eq:dsdo}), reproduce the cross sections and their uncertainties
given in Table~I of Ref.~\cite{Ass06}, for both the two-parameter and
three-parameter approaches described therein.
The phase $\phi$ and its error was also well reproduced in the
case of the three-parameter fit.
The fit was determined by $\chi^2$ minimization, with the parameter errors
indicating the range of parameter values consistent with an increase
in $\chi^2$ by an amount $\le1$ above the minimum
when the remaining free parameters are varied.
The uncertainty in $\tilde{N}$
is propagated after fitting the angular distributions.
The uncertainty in center-of-mass (CM) energy, given in column~2 of
Table~I of Ref.~\cite{Ass06}, is propagated to the $S$ factors when
calculating them from the cross section.

These fits also describe the reported angular distribution data via
$N_\gamma(\theta)=N_{\gamma 0} W(\theta)$.
We have removed the correction for finite solid from the data by the
transformation $N_\gamma(\theta)\rightarrow
N_\gamma(\theta)+N_{\gamma 0}[W_Q(\theta)-W(\theta)]$, where $N_{\gamma 0}$
and $\sigma_{E1}/\sigma_{E2}$ are taken from the two-parameter fit.
This approach by construction ensures that the identical results will
be produced when the modified $N_\gamma(\theta)$ is fitted
with the attenuation coefficients taken into account.
This step was taken because, in the simultaneous fits of multiple
angular distributions, we feel it is not desirable to have already made
corrections that are based on single-energy fits.
It also makes it possible to investigate the effects of uncertainties in
the $Q_l$. These factors are given in Ref.~\cite{Ass06},
having been calculated from a simple formula, their Eq.~(4.2).
We have calculated the $Q_l$ independently, using a {\sc geant4}
simulation~\cite{Ago03}, taking into account the tapered geometry of the
EUROGAM detectors~\cite{Nol90, Bea92} and the angle-dependent detector
distances (see Table~3.2 of Ref.~\cite{Fey04}).
The changes in the extracted cross sections from using our $Q_l$
were generally found to be negligible.
The one exception was for the four $E1$ cross sections underlying
the narrow $E2$ resonance (measurements 18-21).
However, these four $E1$ cross sections also have very large statistical
uncertainties, rendering the changes to be of less importance.
We have thus used the $Q_l$ factors given in Ref.~\cite{Ass06} in
all subsequent analysis.

Although the ``effective energy'' is extensively discussed by \citet{Ass06},
Refs.~\cite{Ass06,Fey04} do not explicitly state how it is defined.
They presumably used what is called in this work the median energy, as that
was utilized in a slightly earlier analysis by the same group;
see Eqs.~(3.19) and~(3.20) of Ref.~\cite{Kun02}.

\subsection{Center of Mass Motion}

The analysis given in Refs.~\cite{Ass06,Fey04} did not consider the
effects of CM motion. We have included it in our analysis,
using Eq.~(B9) given by \citet{Bru01}. This approach leads to
a modified form for the angular distribution:
\begin{equation}
W_{Q\beta}(\theta)=P_{Q\beta 0}+\sum_{l=1}^4 a_l P_{Q\beta l}(\cos\theta),
\label{eq:dsdo_w_q_beta}
\end{equation}
where
\begin{eqnarray}
P_{Q\beta 0} &\equiv& 1+2\beta Q_1 P_1, \\ \nonumber
P_{Q\beta l} &\equiv& Q_l P_l+\frac{\beta}{2l+1}
  [(l+1)(l+2)Q_{l+1}P_{l+1}-l(l-1)Q_{l-1} P_{l-1}],
\end{eqnarray}
and $\beta$ is the speed of the recoiling ${}^{16}{\rm O}$ nucleus
relative to the speed of light.
We have utilized a value of 0.849 for the required
$Q_5$ attenuation coefficient.
The inclusion of these effects makes some non-negligible changes,
particularly for the extracted $E2$ cross sections with $E_{c.m.}$ between
2.0 and 2.6~MeV, which are reduced by 10-15\%.
The reduction is very similar to the case reported in Ref.~\cite{Bru01}.
The effects of CM motion have been included for
all analysis described below, with $\beta$
calculated for the median energies in the target reported by
Ref.~\cite{Ass06}. We have verified that this approximation
(ignoring the energy dependence of $\beta$ over the beam-energy loss in
the target) is excellent.
It should be noted that the resulting formula for the differential
cross section remains a linear combination of three terms proportional to
$\sigma_{E1}$, $\sigma_{E2}$, and $\cos\phi\sqrt{\sigma_{E1}\sigma_{E2}}$.
With the assumption of constant $\beta$, all of the energy dependence
of the differential cross section is contained in these three quantities.

\subsection{Energy Convolution Model}

The targets consisted of ${}^{12}{\rm C}$ implanted into gold.
The depth distribution can be described by $g(y)$, where $g$ is
the ratio of ${}^{12}{\rm C}$ to $({}^{12}{\rm C}+{\rm Au})$ and
$y$ measures the depth in units of $({}^{12}{\rm C}+{\rm Au})$ atoms/cm${}^2$.
Equation~\ref{eq:y_thick} becomes
\begin{equation}
Y(E_\alpha) = \int_0^{y_0} \sigma(E(y))\,g(y)\,{\rm d}y,
\end{equation}
where $E_\alpha$ is the incident $\alpha$ energy,
$y_0$ is the maximum depth considered (where $g(y)\rightarrow 0$),
and the variation of energy with depth is determined by solving
the differential equation
\begin{equation}
\frac{{\rm d}E}{{\rm d}y}(y)=
  -g(y)\,\epsilon_{\rm C}(E)-[1-g(y)]\,\epsilon_{\rm Au}(E),
\end{equation}
subject to $E(0)=E_\alpha$,
where $\epsilon_{\rm C}$ and $\epsilon_{\rm Au}$ are the stopping powers for
helium ions in carbon and gold, respectively~\cite{Zie08}.
The total number of carbon atoms per unit area in the distribution
$g(y)$ is given by
\begin{equation}
\tilde{N}_g = \int_0^{y_0} g(y)\,{\rm d}y.
\end{equation}
We have neglected energy straggling, as investigations
into this process indicated that this effect is small compared
to changes associated with uncertainties in $g(y)$.
The calculated target-averaged differential cross section
can thus be defined to be
\begin{equation}
\left\langle\frac{{\rm d}\sigma}{{\rm d}\Omega}\right\rangle=
  \frac{1}{\tilde{N}_g}
  \int_0^{y_0} \frac{{\rm d}\sigma}{{\rm d}\Omega}(E)\,g(y)\,{\rm d}y;
\label{eq:dsdo_tgt}
\end{equation}
recall that the corresponding experimental value was noted
in Subsec.~\ref{subsec:fey_data} to be
\begin{equation}
\left\langle\frac{{\rm d}\sigma}{{\rm d}\Omega}\right\rangle_{\rm exp}=
\frac{N_\gamma(\theta)}{0.624(4\pi)Q\tilde{N}}.
\label{eq:exp_tgt_ave}
\end{equation}
We have thus allowed for the possibility that $g(y)$
has a normalization that does not correspond to the experimentally-determined
areal density of carbon, i.e., that $\tilde{N}\neq\tilde{N}_g$.

\subsection{Three-Parameter Fits Near the Narrow $E2$ Resonance}
\label{subsec:narrow}

\begin{figure}[tbh]
\begin{center}
\includegraphics[width=4in]{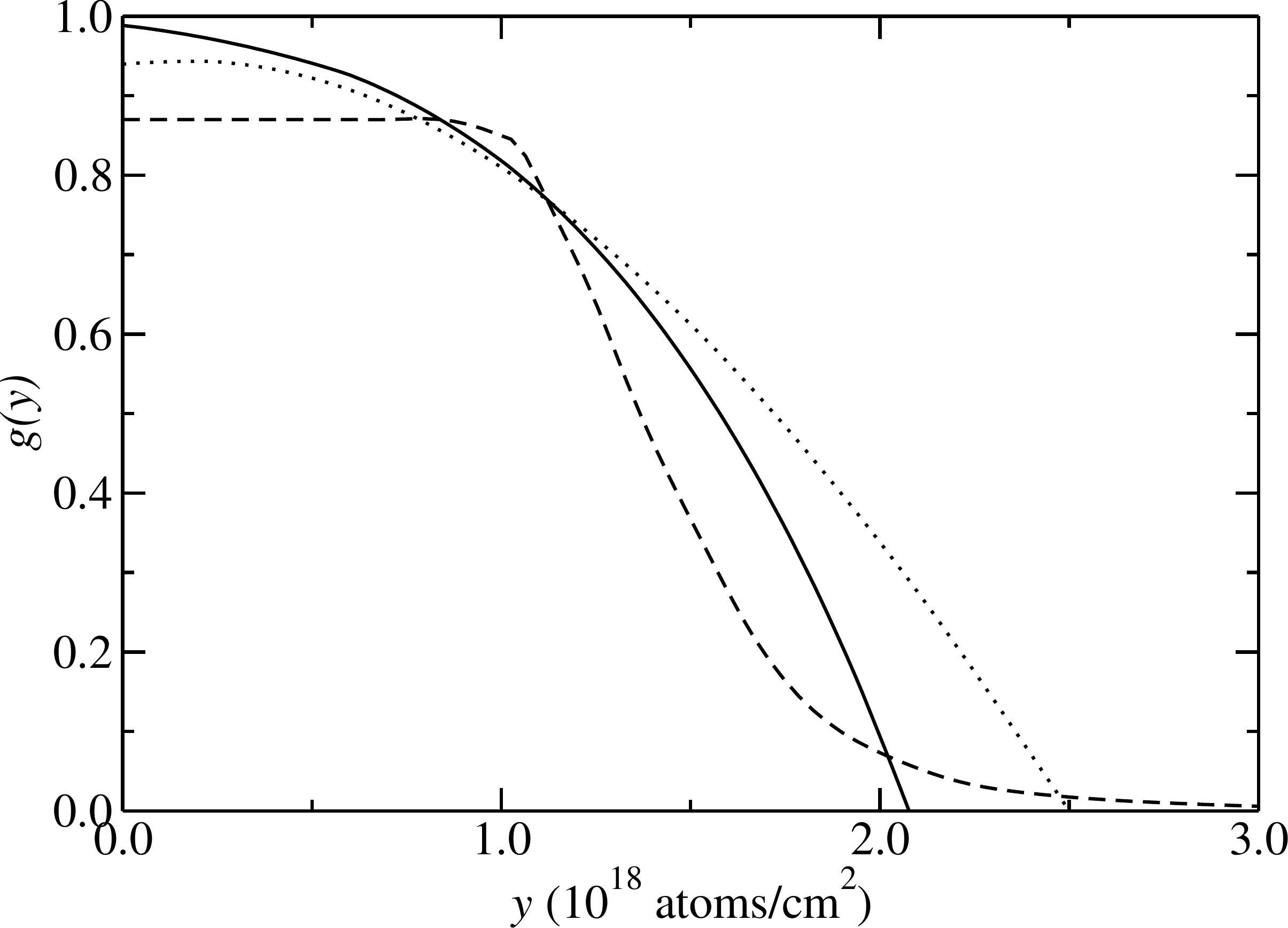}
\caption{Depth profiles $g(y)$ for the targets used near
the narrow $E2$ resonance.
The before (solid curve) and after (dotted curve) profiles are taken from
Fig.~8 of Ref.~\protect\cite{Ass06}. The dashed curve is discussed
in the text.}
\label{fig:depth-profile-1}
\end{center}
\end{figure}

We begin by considering measurements 13-25 which are near the narrow
$E2$ resonance and were  conducted with very similar targets.
For bombarding energies at and above the resonance energy, the effects
of energy convolution are substantial.
At this stage, we will proceed by comparing measurements of the
target-averaged cross sections to calculated convolutions,
as opposed to proceeding straight to deconvolutions.

\begin{figure}[tbhp]
\begin{center}
\includegraphics[width=4in]{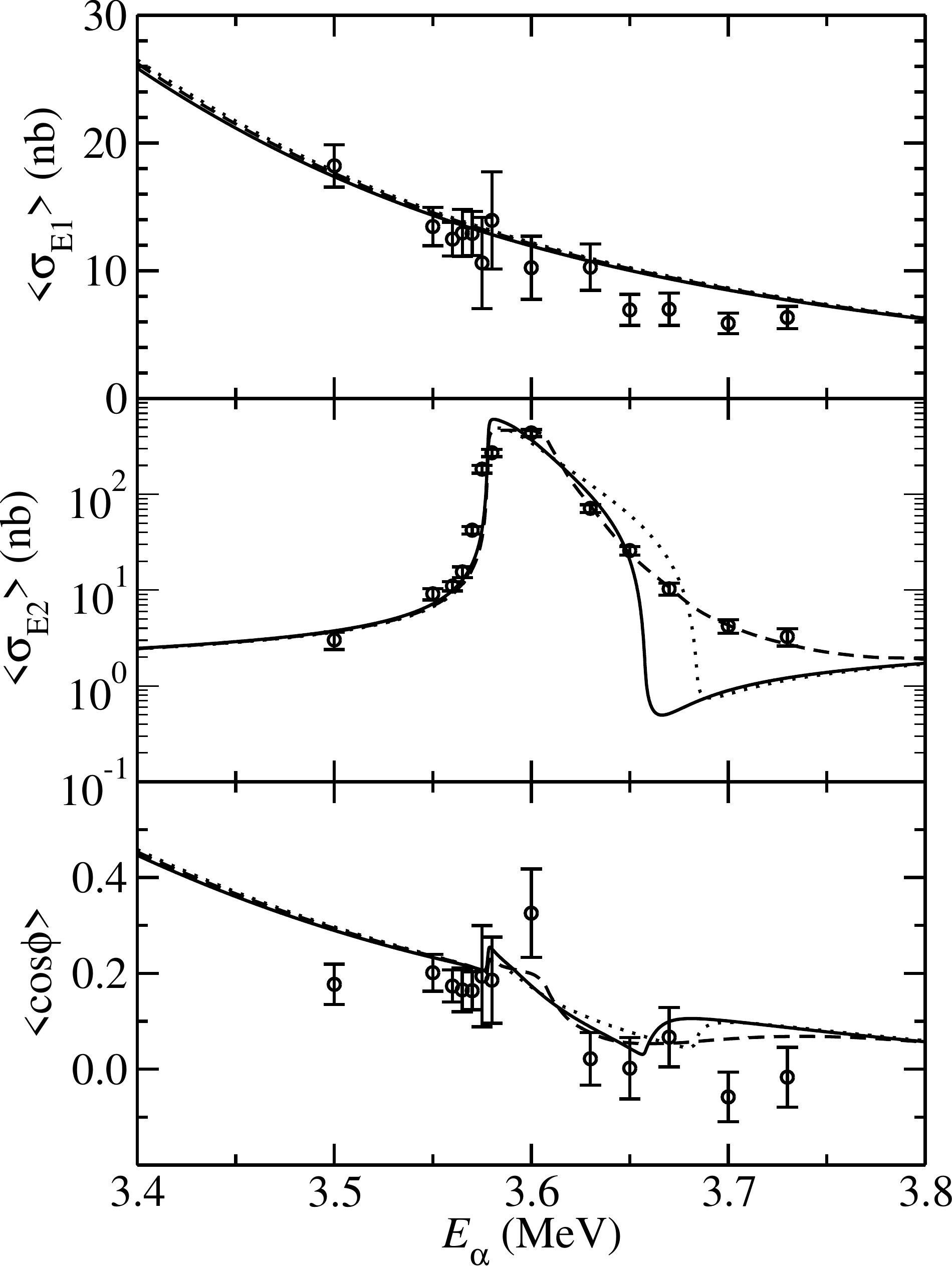}
\caption{Results for the experimental and theoretical
$\langle\sigma_{EL}\rangle$ and $\langle\cos\phi\rangle$ versus
bombarding energy.
The results from measurements 13-25 are shown as the circles and
the theoretical calculations are shown as the curves, which
correspond to the depth profiles shown in Fig.~\ref{fig:depth-profile-1}.}
\label{fig:fey-yield}
\end{center}
\end{figure}

A three-parameter fit to the experimental target-averaged
differential cross sections may now be performed using
\begin{eqnarray}
\left\langle\frac{{\rm d}\sigma}{{\rm d}\Omega}\right\rangle_{\rm exp} &=&
  \frac{1}{4\pi}\biggl[ \langle\sigma_{E1}\rangle(P_{Q\beta 0}-P_{Q\beta 2})
  +\langle\sigma_{E2}\rangle(P_{Q\beta 0}+\frac{5}{7}P_{Q\beta 2}
  -\frac{12}{7}P_{Q\beta 4}) \nonumber \\
&&+6\langle\cos\phi\rangle\sqrt{\frac{\langle\sigma_{E1}\rangle
  \langle\sigma_{E2}\rangle}{5}}
  (P_{Q\beta 1}-P_{Q\beta 3})\biggr],
\label{eq:target_ave_three}
\end{eqnarray}
to yield experimental values for $\langle\sigma_{E1}\rangle$,
$\langle\sigma_{E2}\rangle$, and $\langle\cos\phi\rangle$.
The theoretical expectations for these quantities are:
\begin{eqnarray}
\langle\sigma_{EL}\rangle &=& \frac{1}{\tilde{N}_g}
  \int_0^{y_0}\sigma_{EL}\,g(y)\,{\rm d}y \\
\langle\cos\phi\rangle &=&
\frac{1}{\tilde{N}_g\sqrt{\langle\sigma_{E1}\rangle\langle\sigma_{E2}\rangle}}
  \int_0^{y_0}\cos\phi\sqrt{\sigma_{E1}\sigma_{E2}}\,g(y)\,{\rm d}y,
\end{eqnarray}
where $L=1,2$ and the energy dependences of the cross sections and
$\cos\phi$ under the integrals have been suppressed.

These measurements were carried out with targets containing
approximately $1.3\times 10^{18}$ ${}^{12}{\rm C}$/cm${}^2$.
Example before-and-after depth profiles of such a target are given in Fig.~8
of Ref.~\cite{Ass06} and are also shown in Fig.~\ref{fig:depth-profile-1}.
The results for the experimental and theoretical
$\langle\sigma_{EL}\rangle$ and $\langle\cos\phi\rangle$
are shown in Fig.~\ref{fig:fey-yield}.
In addition to the two depth profiles from Ref.~\cite{Ass06}, we have
considered a third profile that was adjusted to optimize the agreement
with the experimental $\langle\sigma_{E2}\rangle$.
We find that $\langle\sigma_{E1}\rangle$ is insensitive to the details
of the depth profile for all energies considered.
Below the resonance energy, $\langle\sigma_{E2}\rangle$ and
$\langle\cos\phi\rangle$ are also insensitive to the details
of the depth profile. However, at and above the resonance energy,
$\langle\cos\phi\rangle$ and particularly $\langle\sigma_{E2}\rangle$
are very sensitive to the depth profile.

The third depth profile, shown by the dashed curves in
Figs.~\ref{fig:depth-profile-1} and~\ref{fig:fey-yield}, includes a small
tail which extends beyond a depth of $2.5\times 10^{18}$ atoms/cm${}^2$.
It is this tail which allows the calculation to reproduce the two
highest-energy data points.
Note that the two depth profiles from Ref.~\cite{Ass06} do not allow
the resonance to contribute to the $\langle\sigma_{E2}\rangle$
calculation at the two highest energies.
We believe that is extremely likely that $E2$ contribution for these data
points is in fact coming from the resonance and the tail of the depth profile.
The alternative explanation would be that our model cross section is too
low at these energy by a factor of $\sim 5$, which is very unlikely
(see Ref.~\cite{Say12}). In addition, the Rutherford backscattering
technique, which was used to determine the depth profiles in Ref.~\cite{Ass06},
is not sensitive to a small tail.
Further evidence for this interpretation is provided by the gamma-ray
spectra given in Figs.~C.24 and~C.25 of Ref.~\cite{Fey04} that
correspond to these two data points.
The full-energy peaks from the ${}^{12}{\rm C}(\alpha,\gamma_0){}^{16}{\rm O}$
reaction at angles away from $90^\circ$ are double-peaked.
The lower-energy portions of the peaks match the peak energies for the
on-resonance measurements, an observation consistent with significant
reaction yield coming from the resonance.

We conclude that meaningful $E2$ cross section cannot be extracted
for energies at and above the resonance energy ($E_\alpha=3.580$~MeV
and higher) due to the strong sensitivity to the depth profile.
Similar conclusions were reached by \citet{Ass06}, except that
they extracted cross sections and $S$ factors for the two
highest-energy measurements.
For the reasons given in the previous paragraph, we do not believe
these two points can be reliably analyzed.
The $E1$ cross sections can separated from the uncertain $E2$
contribution using the above three-parameter fit procedure;
the $\langle\sigma_{E1}\rangle$ will be analyzed further below.

\subsection{Deconvolution of the Angular Distributions}
\label{subsec:decon_angular}

\begin{figure}[tbh]
\begin{center}
\includegraphics[width=4in]{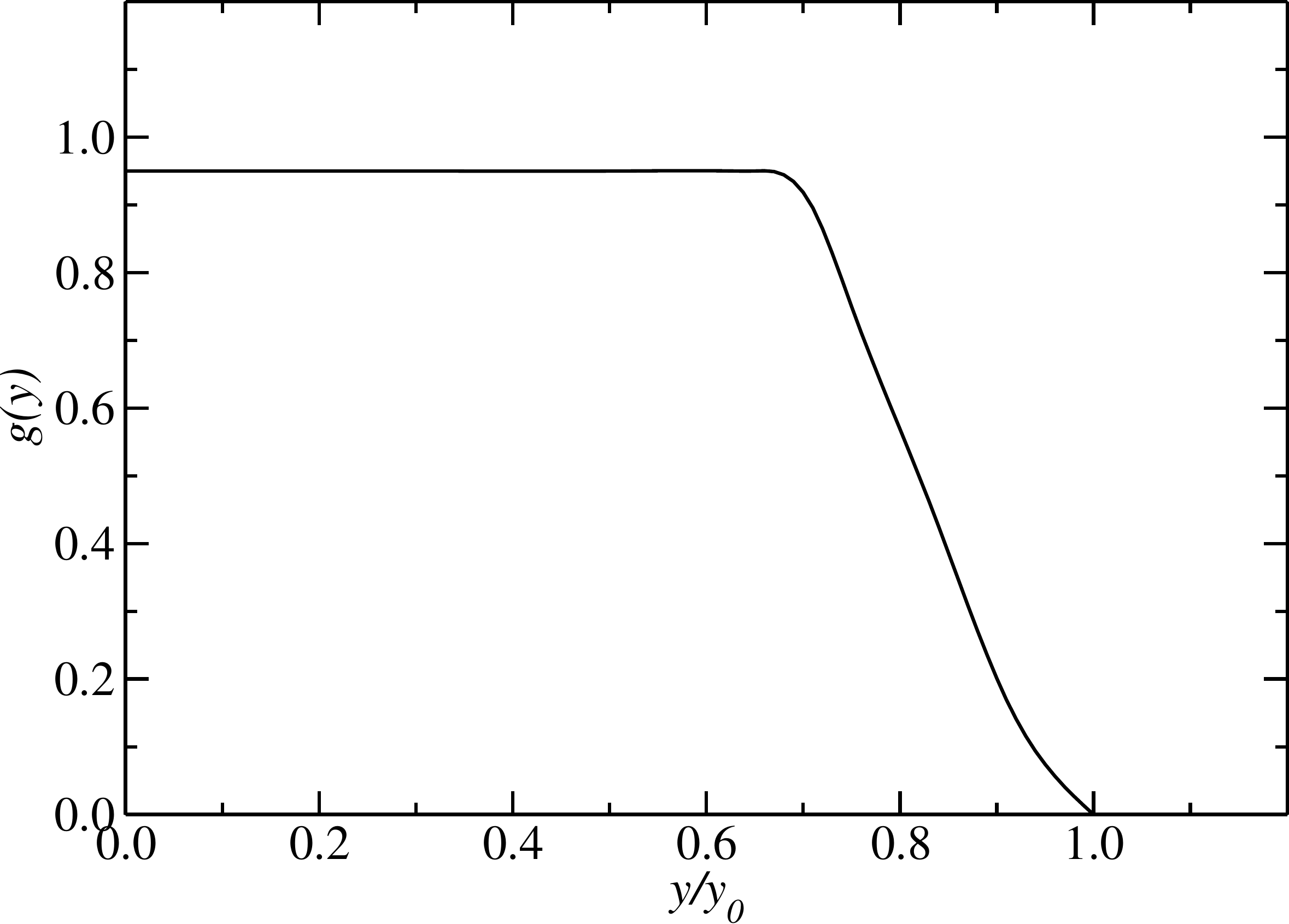}
\caption{The depth profile $g(y)$ for measurements 1-10.}
\label{fig:depth-profile-2}
\end{center}
\end{figure}

The depth profiles utilized in the analysis were defined as follows.
For measurements 1-10, which utilized relatively thick targets,
we have assumed the depth profile shown in Fig.~\ref{fig:depth-profile-2}.
In Ref.~\cite{Ass06}, median energies were determined using the assumption
of a constant $S$ factor.
We have determined the over all scale of these depth profiles by adjusting
the maximum depth $y_0$ to reproduce the the median energies given by
Ref.~\cite{Ass06}, where we have, for this part of the calculation only,
assumed a constant $S$ factor in order to reproduce
the methods of Ref.~\cite{Ass06}.
It should be noted that this procedure generates
depth profiles with ${}^{12}{\rm C}$ areal densities that agree with
those given in Table~II of Ref.~\cite{Ass06} to better than 10\%.
For measurements 11-25, which utilized relatively thinner targets,
we have assumed the depth profile shown by the dashed curve in
Fig.~\ref{fig:depth-profile-1}. In these cases, the depth scale of each
profile has been scaled by $\tilde{N}/1.30$ so that it reproduces the
experimental carbon areal density.

\begin{figure}[tbh]
\begin{center}
\includegraphics[width=4in]{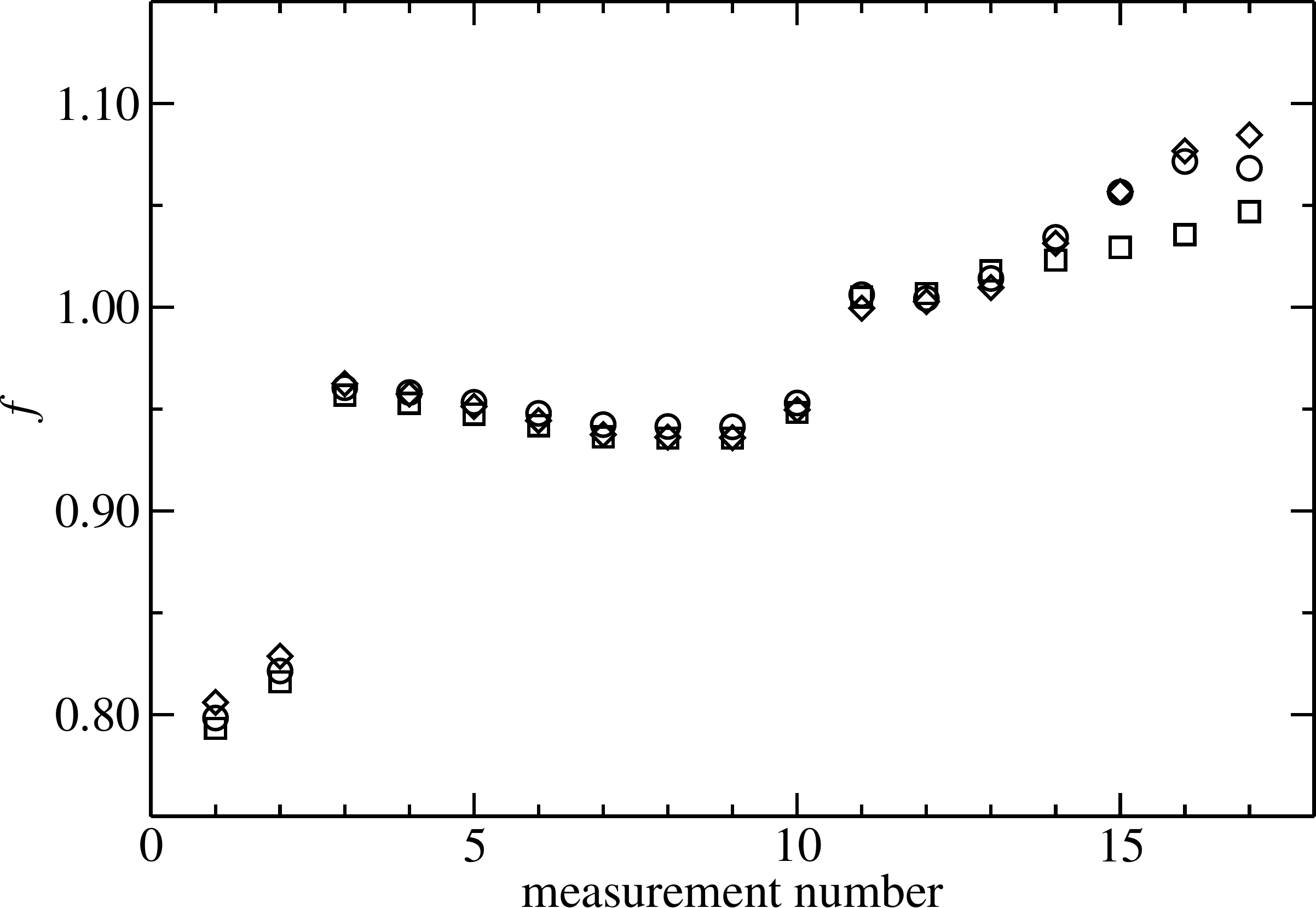}
\caption{The calculated correction factors for $48^\circ$ (circles)
$90^\circ$ (squares), and $130^\circ$ (diamonds) versus measurement number,
for measurements 1-17.}
\label{fig:f-dist-cut1}
\end{center}
\end{figure}

\begin{figure}[tbh]
\begin{center}
\includegraphics[width=4in]{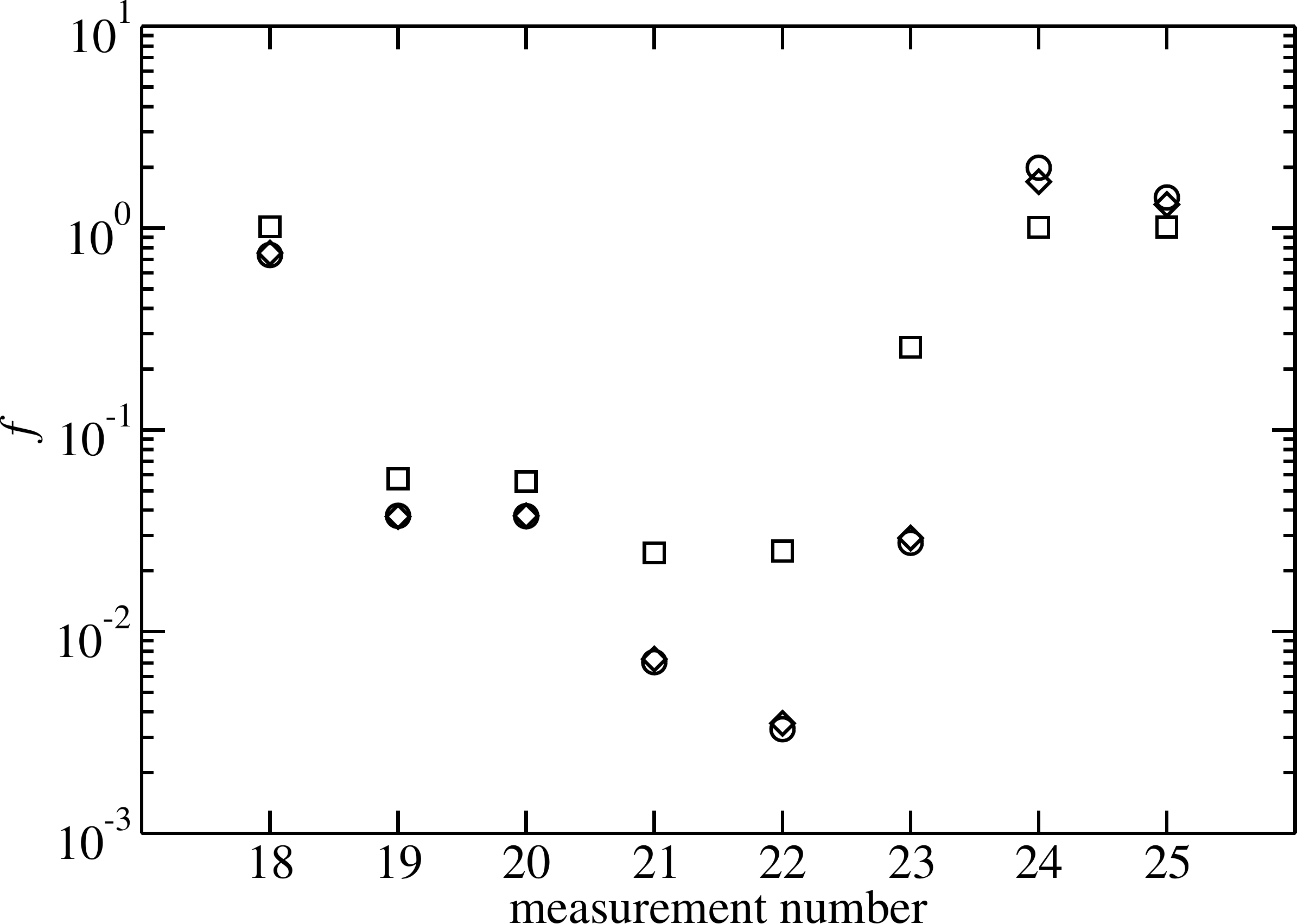}
\caption{The calculated correction factors for $48^\circ$ (circles)
$90^\circ$ (squares), and $130^\circ$ (diamonds) versus measurement number,
for measurements 18-25.}
\label{fig:f-dist-cut2}
\end{center}
\end{figure}

In order to minimize the number of changes in the analysis relative
to Ref.~\cite{Ass06}, the angular distributions will be referenced
to the median energy. The depth of the median energy is defined by
\begin{equation}
\int_0^{y_0} \sigma_0(E(y))\,g(y)\,{\rm d}y =
2\int_0^{y_m} \sigma_0(E(y))\,g(y)\,{\rm d}y,
\end{equation}
where $y_m$ is the median depth and the median energy is given
by $E_m=E(y_m)$.
Note also that we have utilized the total ground state cross section
in the definition.
The deconvoluted differential cross sections, referenced to the median energies
defined above, are calculated from the
experimental target-averaged differential cross sections
(defined by Eq.~\ref{eq:exp_tgt_ave}) via
\begin{equation}
\left(\frac{{\rm d}\sigma}{{\rm d}\Omega}\right)_{\rm exp}=\frac{1}{f}
\left\langle\frac{{\rm d}\sigma}{{\rm d}\Omega}\right\rangle_{\rm exp},
\end{equation}
where the correction factor is given by
\begin{equation}
f=[ \tilde{N}_g \frac{{\rm d}\sigma}{{\rm d}\Omega}(E_m) ]^{-1}
  \int_0^{y_0} \frac{{\rm d}\sigma}{{\rm d}\Omega}(E)\,g(y)\,{\rm d}y.
\label{eq:f_dsdo_fey}
\end{equation}
The theoretical form assumed for the differential
cross section is the same as used in Subsec.~\ref{subsec:narrow}:
\begin{equation}
\frac{{\rm d}\sigma}{{\rm d}\Omega}(E)=\frac{\sigma_0}{4\pi}W_{Q\beta}(\theta).
\end{equation}
The resulting correction factors are shown in Figs.~\ref{fig:f-dist-cut1}
and~\ref{fig:f-dist-cut2} for three representative angles.
For measurements 1-17 (below the peak of the narrow $E2$ resonance),
the correction factors are seen to differ from unity by at most 20\% and to
depend very little upon the angle.
For measurements 18-25, the correction factors show much great departures from
unity and significant angular dependence; these correction factors are also
very sensitive to the assumed target depth profile.

At this point, it is instructive to investigate how sensitive the deconvolution
is to the assumed form of the cross section and depth profiles, both
through the calculation of the median energy and the correction factor $f$.
We will limit our consideration here to measurements 1-13, as
measurements 14-17 are on the rising slope of the narrow $E2$ resonance
where taking the exact energy dependence of this resonance into account
is clearly important, and it has already been noted that the measurements
18-25 are very sensitive to the assumed depth profile.
To test the sensitivity to the assumed cross section, we have repeated the
deconvolution assuming a constant $S$-factor and $\cos\phi$.
In the case of the median energy, only measurements 1-2 show
any sensitivity: the calculation with our model cross section yields median
energies about 7-keV lower than the constant $S$-factor calculation;
note that this lowering corresponds to a a 4\% increase in the
calculated experimental $S$ factors.
The differences in the calculated correction factors are at most 5\%.
We have also investigated the effects of a 20\% increase in the scale of
the depth profiles. This change leads to reductions in the median energy
of up to 12~keV and up to 9\% reductions in the correction factor.
All these changes are within the tolerance implied by the uncertainties
assigned to the effective energies given in column~2 of Table~II of
Ref.~\cite{Ass06}.

\begin{figure}[tbhp]
\begin{center}
\includegraphics[width=4in]{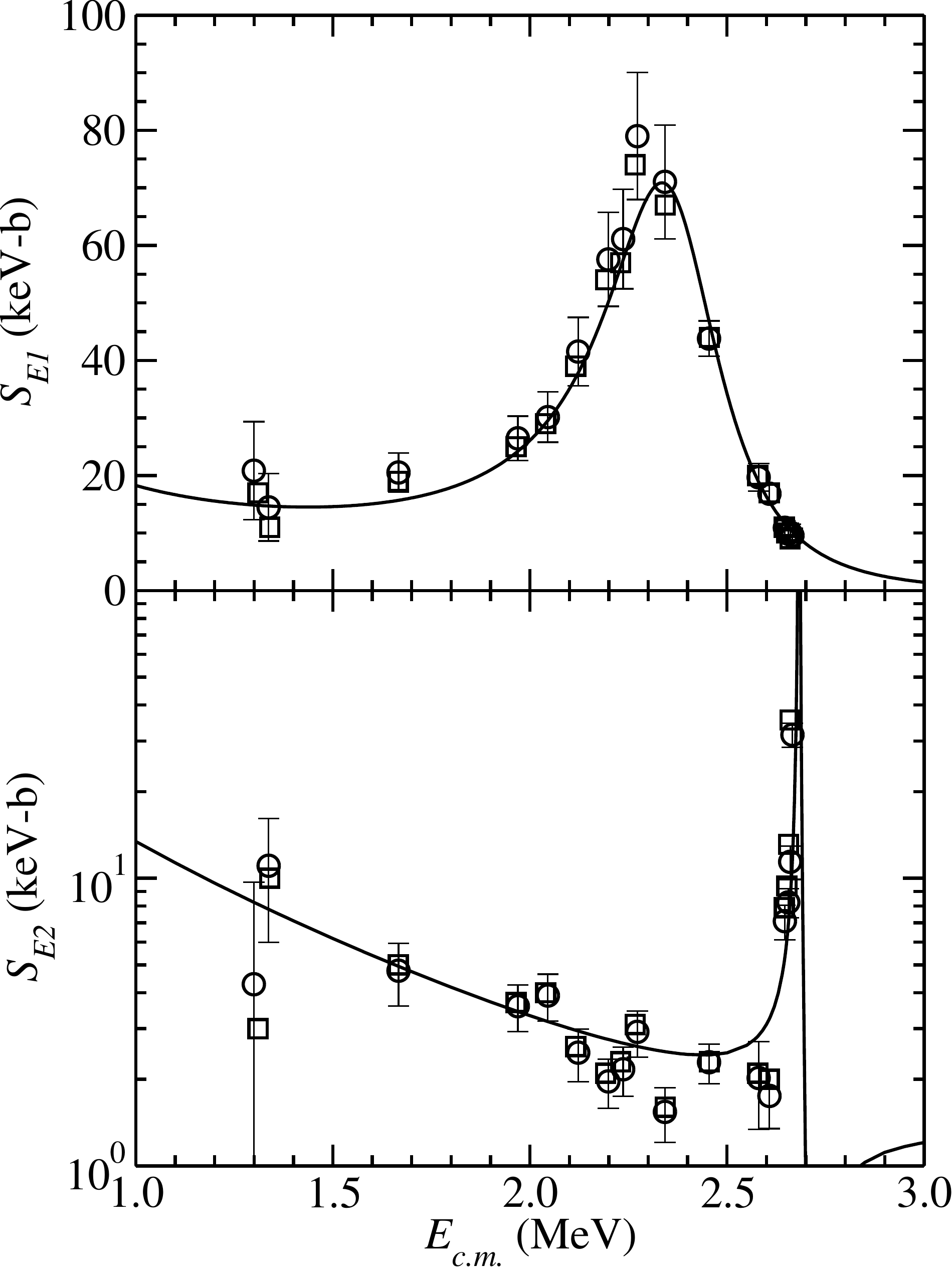}
\caption{The $E1$ (upper panel) and $E2$ (lower panel) $S$ factors.
The values determined from the deconvoluted differential cross sections
for measurements 1-17 are given as circles.
The corresponding values given by Ref.~\cite{Ass06} are given
as squares; the error bars for these points are not shown but are
essentially identical to the error bars on the corresponding circles.
The $R$-matrix parametrizations are given by the solid curves.}
\label{fig:euro22_s-cut-1}
\end{center}
\end{figure}

\begin{figure}[tbhp]
\begin{center}
\includegraphics[width=4in]{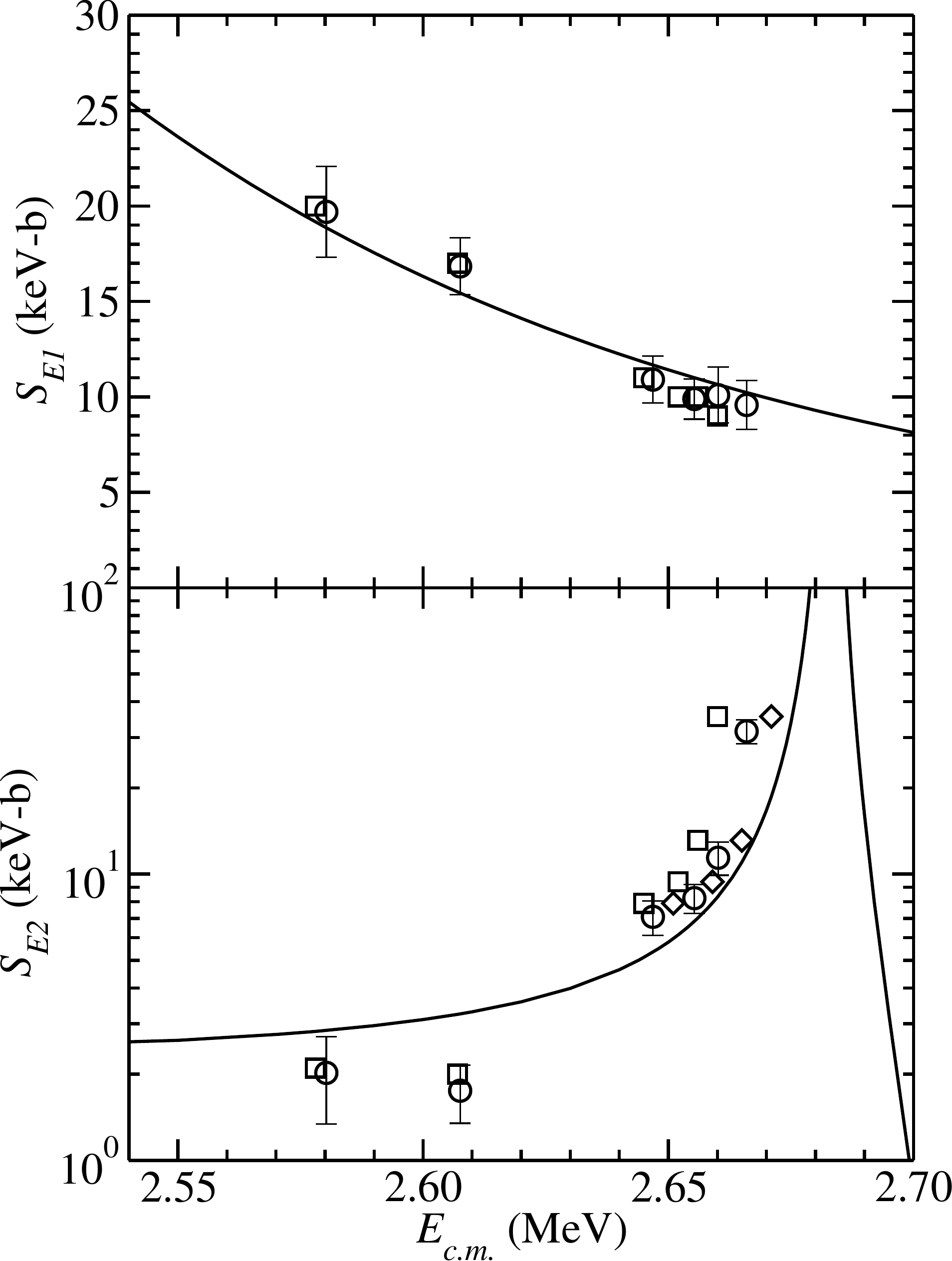}
\caption{The $E1$ (upper panel) and $E2$ (lower panel) $S$ factors.
The values determined from the deconvoluted differential cross sections
for measurements 12-17 are given as circles.
The corresponding values given by Ref.~\cite{Ass06} using their
energy method~(I) are given as squares and for measurements 14-17 using
their energy method~(II) are as diamonds (shown only for the lower panel);
the error bars for these points are not shown but are
essentially identical to the error bars on the corresponding circles.
The respective $R$-matrix parametrizations are given by the solid curves.}
\label{fig:euro22_s-cut-2}
\end{center}
\end{figure}

The deconvoluted experimental differential cross sections for
measurements 1-17 have been fitted using
\begin{eqnarray}
\left(\frac{{\rm d}\sigma}{{\rm d}\Omega}\right)_{\rm exp} &=&
  \frac{1}{4\pi}\biggl[\sigma_{E1}(P_{Q\beta 0}-P_{Q\beta 2})
  +\sigma_{E2}(P_{Q\beta 0}+\frac{5}{7}P_{Q\beta 2}
  -\frac{12}{7}P_{Q\beta 4}) \nonumber \\
&&+6\cos\phi\sqrt{\frac{\sigma_{E1}\sigma_{E2}}{5}}
  (P_{Q\beta 1}-P_{Q\beta 3})\biggr],
\end{eqnarray}
to determine $\sigma_{E1}$ and $\sigma_{E2}$; $\cos\phi$ has
been fixed at the value given by the $R$-matrix parametrization
shown in Fig.~\ref{fig:rmatrix}.
The fit was determined by $\chi^2$ minimization, with the parameter errors
indicating the range of parameter values consistent with an increase
in $\chi^2$ by an amount $\le1$ above the minimum
when the remaining free parameters are varied.
The uncertainties in $\tilde{N}$ and the CM energy
have been propagated after fitting.
The results are shown as $S$-factors in Figs.~\ref{fig:euro22_s-cut-1}
and~\ref{fig:euro22_s-cut-2} and Table~\ref{tab:results}.
For reference, the figures also show results given by Ref.~\cite{Ass06}
from the analysis of the same raw data and the $R$-matrix parametrization
shown in Fig.~\ref{fig:rmatrix}.
Our analysis differs from that given in Ref.~\cite{Ass06} in
three ways: the effects of CM motion have been taken into
account, the median energy has been calculated using a more accurate
representation of the energy dependence of the cross section, and
the differential cross sections have been adjusted by the the
correction factor given by Eq.~\ref{eq:f_dsdo_fey}.
Another difference is that we are fitting differential cross sections
that have not been corrected for detector solid angle effects but rather
include these corrections in our fitting function; this difference should
not change the extracted $S$ factors.
The differences between our results and those given by Ref.~\cite{Ass06}
are generally quite small. In the case of the $E2$ $S$ factor, the
effects of CM motion and the deconvolution tend to cancel,
leading to very small changes.
Our results are also in generally good agreement with the $R$-matrix
parametrization, which indicates consistency between the data and
our cross section model.
Near the narrow $E2$ resonance, Ref.~\cite{Ass06} provided two values
of the energy (columns 2 and~3 of Table~II of that reference); both
values are plotted in Fig.~\ref{fig:euro22_s-cut-2}.
We note that for measurements 14-17, both our results and those of
Ref.~\cite{Ass06} lie 3-10~keV to the left of the $R$-matrix parametrization.
This discrepancy is also seen in Fig.~\ref{fig:fey-yield}.
The $R$-matrix parametrizaton should be well determined here by the known
parameters of the resonance.
One possibility is that uncertainty in the beam energy contributes to this
difference.
In any case, the deconvolution procedure for measurements 14-17 must be
viewed with some skepticism due to the lack of agreement between the resulting
data points and our cross section model.

\begin{figure}[tbhp]
\begin{center}
\includegraphics[width=4in]{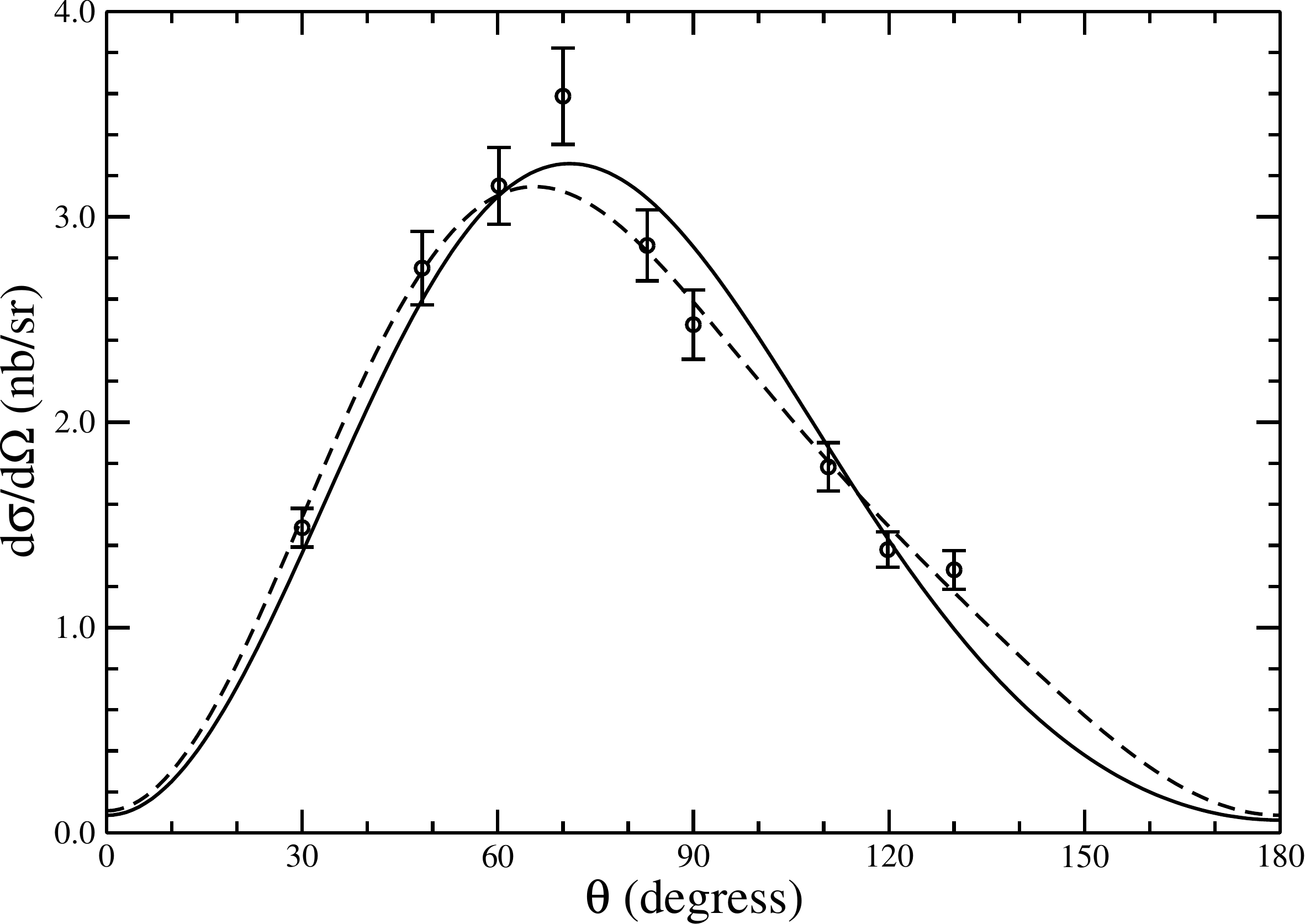}
\caption{The deconvoluted angular distribution data for measurement~7.
The solid and dashed curves show the two-parameter and three-parameter
fits that are discussed in more detail in the text.}
\label{fig:dsdo_7}
\end{center}
\end{figure}

In Fig.~\ref{fig:dsdo_7}, we show the differential cross section obtained
for measurement~7, for which we find median energy to be
$E=2.199$~MeV in the CM system.
The two-parameter fit is shown as the solid curve; this fit utilized
a fixed value of $\cos\phi=0.941$ and yields
$\sigma_{E1}=24.3(5)$~nb and $\sigma_{E2}=0.8(1)$~nb with $\chi^2=22.0$.
\citet{Ass06} noted that three-parameter fits, where $\cos\phi$
was allowed to vary, tended to differ significantly from two-parameter
fits for CM energies between~2.0 and~2.4~MeV.
We find similar results.
Such a three parameter fit for measurement~7 is also shown in
Fig.~\ref{fig:dsdo_7} as the dashed curve; it yields
$\sigma_{E1}=21.9(8)$~nb, $\sigma_{E2}=3.2(6)$~nb, and
$\cos\phi=0.49(6)$ with $\chi^2=8.3$.
The above uncertainties do not include the contribution from
$\tilde{N}$ which is common to all angles of a single angular distribution.
It is seen that while both fits give essentially identical total
ground-state cross sections ($\sigma_0=\sigma_{E1}+\sigma_{E2}$),
the three-parameter fit gives a much larger $\sigma_{E2}$ --
by a factor of 3.8 which corresponds to 3.6 standard deviations.
We would like to emphasize that there is no known reason for the
value for $\cos\phi$ assumed in the two-parameter fit to be
incorrect (see discussion in Ref.~\cite{Bru01}).
We have investigated the possible role of uncertainties in
the deconvolution procedure on this finding.
We conclude that it cannot be explained by this consideration, due to
the lack of angle dependence of the convolution factor in this
energy range (see Fig.~\ref{fig:f-dist-cut1}).
The most likely explanation is that angular distribution data
contain unidentified systematic errors.
Evidence for this explanation is provided by the $\chi^2$ values
from both the two-parameter and three-parameter fits to the data.
Our $\chi^2$ values are very similar to those of Ref.~\cite{Ass06};
as can be seen from Table~I of Ref.~\cite{Ass06}, the $\chi^2$ values
always exceed the number of degrees of freedom and often do so by
a substantial amount. Another observation is that particular angles
show systematic deviations from the fits.
For example, the $130^\circ$ data point lies above the fitted curve
for 23 out of the 25 distributions in the case of the three-parameter fits
(data near the narrow $E2$ resonance discussed in Subsec.~\ref{subsec:narrow}
were also considered).

\begin{table}[tbhp]
\begin{tabular}{r|cccc|cc}
   & $E_{\rm c.m.}$ & $S_{E1}$ & $S_{E2}$ & $S_0$  &
  $E_{c.m.}$ & $S_{E1}$ \\
   & (MeV)          & (keV-b)  & (keV-b)  & (keV-b)&
  (MeV)      & (keV-b) \\ \hline
 1 &  1.299 & 20.8(85)  &  4.3(54)  & 25.1(73)  &        &          \\
 2 &  1.337 & 14.5(59)  & 11.0(51)  & 25.5(69)  &        &          \\
 3 &  1.667 & 20.5(34)  &  4.8(12)  & 25.3(39)  &        &          \\
 4 &  1.969 & 26.5(39)  &  3.6(7)   & 30.1(43)  &        &          \\
 5 &  2.045 & 30.1(44)  &  3.9(7)   & 34.1(49)  &        &          \\
 6 &  2.123 & 41.5(60)  &  2.5(5)   & 44.0(63)  &        &          \\
 7 &  2.199 & 57.6(82)  &  2.0(4)   & 59.5(84)  &        &          \\
 8 &  2.236 & 61.1(86)  &  2.2(4)   & 63.3(89)  &        &          \\
 9 &  2.272 & 79.0(110) &  2.9(5)   & 81.9(114) &        &          \\
10 &  2.343 & 71.0(99)  &  1.5(3)   & 72.6(101) &        &          \\
11 &  2.454 & 43.8(31)  &  2.3(4)   & 46.1(32)  &        &          \\
12 &  2.580 & 19.7(24)  &  2.0(7)   & 21.7(25)  &        &          \\
13 &  2.608 & 16.8(15)  &  1.8(4)   & 18.6(16)  &        &          \\
14 &  2.647 & 10.9(12)  &  7.1(10)  & 18.0(16)  &  2.645 & 10.9(12) \\
15 &  2.655 &  9.9(11)  &  8.2(10)  & 18.1(16)  &  2.653 & 10.0(11) \\
16 &  2.660 & 10.1(15)  & 11.4(15)  & 21.5(20)  &  2.656 & 10.3(15) \\
17 &  2.666 &  9.6(13)  & 31.5(30)  & 41.1(36)  &  2.660 & 10.2(14) \\
18 &        &           &           &           &  2.664 &  8.3(28) \\
19 &        &           &           &           &  2.668 & 10.8(29) \\
20 &        &           &           &           &  2.683 &  7.7(19) \\
21 &        &           &           &           &  2.706 &  7.4(13) \\
22 &        &           &           &           &  2.721 &  4.8(9)  \\
23 &        &           &           &           &  2.736 &  4.8(9)  \\
24 &        &           &           &           &  2.759 &  3.8(5)  \\
25 &        &           &           &           &  2.781 &  4.0(5)  \\
\end{tabular}
\caption{The deconvoluted $S$-factor data. The first column supplies
the measurement number. The following four columns provide the $E1$,
$E2$, and total $S$-factor data from the two-parameter fits discussed
in Subsec.~\ref{subsec:decon_angular}. The final two columns provide
the $E1$ $S$-factor data from the three-parameter fits discussed in
Subsec.~\ref{subsec:E1}.}
\label{tab:results}
\end{table}

In Table~\ref{tab:results} we also provide the total ground-state
$S$~factors $S_0$, where the uncertainty is determined using
two-parameter fits considering $\sigma_0$ and $\sigma_{E2}/\sigma_{E1}$
to be independent variables.
The uncertainties in $\tilde{N}$ and the CM energy
have again been propagated after fitting.
In a simultaneous $R$-matrix analysis of data from this experiment,
we envision fitting fitting $S_0$ and the angular distributions.
The angular distributions would be considered to be un-normalized,
with the normalization carried in the $S_0$ data, and
an overall normalization uncertainty of 9\% would also be assumed
for the entire data set (inferred from Table~5.2 of Ref.~\cite{Fey04}).
This approach ensures that only independent data with independent uncertainties
are fitted; this would not be the case if both the $S_{E1}$ and
$S_{E2}$ values were included in the fit.

\subsection{Deconvolution of $E1$ data}
\label{subsec:E1}

In Subsec.~\ref{subsec:narrow} it was noted that target-averaged
$E1$ cross sections can be extracted independently of the
$E2$ cross sections. This analysis will be pursued further here, as it
is useful for the data near the narrow $E2$ resonance where the
$E2$ contribution is extremely sensitive to the assumed target profile.
In this case, we define the median energy using the $E1$ cross section:
\begin{equation}
\int_0^{y_0} \sigma_{E1}(E(y))\,g(y)\,{\rm d}y =
2\int_0^{y_m} \sigma_{E1}(E(y))\,g(y)\,{\rm d}y,
\end{equation}
where the median energy is given by $E_m=E(y_m)$.
This procedure avoids introducing un-needed uncertainty through
$E2$ contribution.
The same target profiles are used as in Subsec.~\ref{subsec:decon_angular}.
The target-averaged $E1$ cross sections $\langle\sigma_{E1}\rangle$ are
obtained using three-parameter fits to Eq.~(\ref{eq:target_ave_three}).
Deconvoluted $E1$ cross sections are obtained via
\begin{equation}
(\sigma_{E1})_{\rm exp} = \frac{1}{f_{E1}} \langle\sigma_{E1}\rangle
\end{equation}
where in this case the correction factor is given by
\begin{equation}
f_{E1}=[ \tilde{N}_g \sigma_{E1}(E_m) ]^{-1}
  \int_0^{y_0} \sigma_{E1}(E)\,g(y)\,{\rm d}y.
\end{equation}
The resulting $E1$ $S$~factors for measurements 14-25 are given in
Table~\ref{tab:results}.

\subsection{Electronic Files}

Three supplementary files accompany this paper on the publisher's website.
These files contain the $N_\gamma(\theta)$ data from Ref.~\cite{Fey04},
the $N_\gamma(\theta)$ data with the correction for finite solid angle
removed, and the $N_\gamma(\theta)$ data with the correction for
finite solid angle removed and the deconvolution factor given by
Eq.~(\ref{eq:f_dsdo_fey}) applied, respectively.
The latter data, when normalized by $0.624(4\pi)Q\tilde{N}$,
become the deconvoluted experimental differential cross sections in nb/sr.

\section{Conclusions}

A general framework for deconvoluting the effects of energy averaging
on charged-particle reaction measurements has been presented.
There are many potentially correct approaches to the problem; the
relative merits of some of them have been discussed.

These deconvolution methods have been applied to recent
${}^{12}{\rm C}(\alpha,\gamma){}^{16}{\rm O}$ measurements~\cite{Ass06,Fey04}.
This analysis has clarified what can, and what cannot, be explained by
energy convolution effects.
Very significant effects are found at and above the narrow 2.68-MeV resonance.
Below this resonance,
we find that considerations of energy deconvolution, as well as
special relativity, lead to relatively small changes in the extracted
$S$~factors. We have also extracted deconvoluted differential
cross sections from this experiment.
We expect that these results will be useful for future analyses,
including a simultaneous $R$-matrix of the angular distributions.

\section*{Acknowledgments}

We thank Lothar Buchmann for useful discussions regarding the
${}^{12}{\rm C}(\alpha,\gamma){}^{16}{\rm O}$ reaction.
This work was supported in part by the U.S. Department
of Energy under grant numbers DE-FG02-88ER40387 and DE-FG52-09NA29455.





\bibliographystyle{model1a-num-names}
\bibliography{decon-ctag}







\end{document}